\documentclass[twocolumn]{article}
\pdfoutput=1
\usepackage{titling} 
\usepackage{graphicx}  
\usepackage{booktabs}  
\usepackage{multirow}  
\usepackage{makecell}  
\usepackage{mathtools} 
\usepackage{amssymb}   
\usepackage{siunitx}   
\usepackage{caption}   
\usepackage[hidelinks]{hyperref}  
\usepackage{amsmath}   
\usepackage{braket}    
\usepackage{cleveref}  
\usepackage{placeins}  
\usepackage{authblk}   
\usepackage{arydshln}  
\usepackage{caption}
\usepackage[english]{babel}
\usepackage{stfloats}
\usepackage[super]{cite} 
\usepackage[top=2.5cm, bottom=2.5cm, left=2cm, right=2cm]{geometry}
\usepackage{ragged2e} 
\usepackage{parskip}
\pretitle{\begin{center}\Large\bfseries}
\posttitle{\end{center}\vspace{5mm}}

\newcolumntype{C}[1]{>{\centering\arraybackslash}m{#1}} 

\newcommand{\muEG}{\mu_{eg}}
\newcommand{\muEGdag}{\mu_{eg}^{\dagger}}
\newcommand{\muGE}{\mu_{ge}}
\newcommand{\muGEdag}{\mu_{ge}^{\dagger}}
\newcommand{\UEE}[1]{U_{ee}(#1)}
\newcommand{\UEEdag}[1]{U_{ee}^{\dagger}(#1)}
\newcommand{\UGG}[1]{U_{gg}(#1)}
\newcommand{\UGGdag}[1]{U_{gg}^{\dagger}(#1)}
\newcommand{\wavenumbers}{\ensuremath{\mathrm{cm}^{-1}}}

\title{Using machine learning to map simulated noisy and laser-limited multidimensional spectra to molecular electronic couplings}

\author[1]{Jonathan D. Schultz\thanks{Email: jonathan.schultz@duke.edu}}
\author[1]{Kelsey A. Parker\thanks{Email: kelsey.parker@duke.edu}}
\author[1,2]{Bashir Sbaiti}
\author[1,2,3]{David N. Beratan}

\affil[1]{Department of Chemistry, Duke University, Durham, NC 27708, United States} 
\affil[2]{Department of Physics, Duke University, Durham, NC 27708, United States} 
\affil[3]{Department of Biochemistry, Duke University, Durham, NC 27710, United States}

\date{} 

\begin{document}

\twocolumn[
   \begin{@twocolumnfalse}
        \maketitle
        \vspace{-1em}
        \begin{quotation}\small\noindent
            Two-dimensional electronic spectroscopy (2DES) has enabled significant discoveries in both biological and synthetic energy-transducing systems. Although deriving chemical information from 2DES is a complex task, machine learning (ML) offers exciting opportunities to translate complicated spectroscopic data into physical insight. Recent studies have found that neural networks (NNs) can map simulated multidimensional spectra to molecular-scale properties with high accuracy. However, simulations often do not capture experimental factors that influence real spectra, including noise and suboptimal pulse resonance conditions, bringing into question the experimental utility of NNs trained on simulated data. Here, we show how factors associated with experimental 2D spectral data influence the ability of NNs to map simulated 2DES spectra onto underlying intermolecular electronic couplings. By systematically introducing multisourced noise into a library of 356000 simulated 2D spectra, we show that noise does not hamper NN performance for spectra exceeding threshold signal-to-noise ratios (SNR) ($> 6.6$ if background noise dominates vs. $> 2.5$ for intensity-dependent noise). In stark contrast to human-based analyses of 2DES data, we find that the NN accuracy improves significantly (ca. $84$\% $\rightarrow$ $96$\%) when the data are constrained by the bandwidth and center frequency of the pump pulses. This result is consistent with the NN learning the optical trends described by Kasha's theory of molecular excitons. Our findings convey positive prospects for adapting simulation-trained NNs to extract molecular properties from inherently imperfect experimental 2DES data. More broadly, we propose that machine-learned perspectives of nonlinear spectroscopic data may produce unique and, perhaps counterintuitive guidelines for experimental design.
        \end{quotation}
        \vspace{2em}
   \end{@twocolumnfalse}
]

\section{Introduction}

Coherent multidimensional spectroscopies (CMDS) afford rich insight into the mechanisms of light-driven molecular processes.\cite{biswas2022coherent,fuller2015experimental,collini20212d,scholes2017using,dean2017coherence,fresch2023two, schultz2024coherence} For example, studies using two-dimensional electronic spectroscopy (2DES) in the last two decades exposed the central role that electron-vibrational (vibronic) coupling plays in the excited-state photophysics of chemical and material systems, including natural photosynthetic complexes,\cite{tempelaar2014vibrational,chenu2015coherence,fuller2014vibronic,dean2016vibronic} organic semiconductors,\cite{bakulin2016real,schultz2021influence,de2016tracking,song2014vibrational,de2021intermolecular} and quantum dots.\cite{caram2014exploring,collini2021ultrafast} The abundance of information within 2DES spectra, as with spectra from other CMDS techniques, comes at the expense of interpretability; results of early 2DES measurements sparked decade-long debates of their physical interpretation.\cite{cao2020quantum,duan2017nature,manvcal2020decade,zerah2021photosynthetic, schultz2024coherence} Developing robust methods to derive accurate chemical information from 2DES will be indispensable as this technique is used increasingly to probe complex, device-relevant condensed-phase systems. 

Spectroscopy is often used to  solve \textit{inverse problems}, where physical insight about a chemical system is sought from  spectroscopic data. Machine learning (ML) models are uniquely suited to solve inverse problems,\cite{karniadakis2021physics,sridharan2022modern} and ML has already been applied to many inverse chemical problems in spectroscopy \cite{meza2021applications,fang2021decoding,lansford2020infrared,enders2021functional,namuduri2020machine,kollenz2020unravelling,ren2022machine,rodriguez2019machine, parker2022mapping, sbaiti2025machine,wu2024unraveling,ye2025ai,lemm2024impact,rieger2023understanding,david2023towards,liu2017deep}. For example, Lansford et al.\cite{lansford2020infrared} and Enders et al.\cite{enders2021functional} used ML to extract surface microstructure and functional group information, respectively, from infrared spectra.  Cui et al.\cite{cui2024quantitative} demonstrated a ML method that relates infrared and Raman spectra to the electrocatalytic properties of CO\textsubscript{2} reduction. Despite recent progress in joint ML-spectroscopic approaches, time-evolving nonlinear spectra are vastly more complicated than steady-state linear spectra, and this is especially true for spectra derived from multidimensional methods like 2DES. As a result, few studies\cite{namuduri2020machine,ren2022machine, parker2022mapping, sbaiti2025machine,wu2024unraveling,ye2025ai,rodriguez2019machine} have demonstrated how ML can be used to map the properties of molecular systems directly from their multidimensional spectra. However, innovations enabled by ML applied to linear spectroscopy\cite{lansford2020infrared,enders2021functional,cui2024quantitative,torrisi2020random,rieger2023understanding} and magnetic resonance spectroscopies\cite{jonas2019deep,khosravian2024hamiltonian,lemm2024impact} clearly indicate the potential of using ML to transform the interpretation of complicated spectroscopic data.

The data requirements of ML pose a significant challenge in applying ML to spectroscopy.\cite{namuduri2020machine,sridharan2022modern,jonas2019deep,schuetzke2023validating} There is currently no public repository for experimental 2DES data. Of the experimental data that accompany journal publications, factors such as low molecular diversity, variation in data processing methods, and insufficient sample characterization hinder the prospects for training neural networks (NNs) with purely experimental 2DES datasets. A potentially viable alternative is to use simulated data to train NNs for experimental applications.\cite{lansford2020infrared,schuetzke2021enhancing,wang2020rapid,chen2020review,yang2025monitoring,han2021transfer} Simulated data offer the unique advantages of practically infinite availability and complete knowledge of the underlying physical properties, which enabled several recent studies\cite{namuduri2020machine,ren2022machine,parker2022mapping,sbaiti2025machine} that leverage ML to solve inverse problems with multidimensional spectra. Simulated data are, however, pristine: they do not typically include the influence of experimental features in 2DES spectra, such as noise, finite pulse bandwidths, and imperfect laser-sample resonance conditions.\cite{fuller2015experimental,kearns2017broadband} It remains unknown how such experimental aspects of 2DES might influence the performance of ML-based interpretation tools.

Here, we develop an expansive database of 356000 vibronic dimer 2DES spectra and use it to identify how experimental constraints influence inverse problem solving with a feed-forward NN. When trained and evaluated on pristine simulated data, the NN classifies unseen spectra to one of 33 electronic coupling categories with $\sim84$\% accuracy. By systematically introducing experimental constraints, or ``data pollutants,'' into the spectra and performing repeated training and evaluation, we find how the pollutants influence the testing performance of the NN. We find that the simulation-trained NNs are relatively robust to noise sources that depend on the signal magnitude (e.g., fluctuations in the pump power or beam alignment). We also find that NN performance \textit{increases} significantly (up to $\sim96$\% accuracy) when the effects of pump bandwidth and center frequency are accounted for in the spectral dataset. We find that this counterintuitive result provides fundamental insight into the machine learnability of electronic coupling information in multidimensional optical spectra. Ultimately, our study provides methods to adapt simulation-trained NNs to experimental applications and clarifies potential practical limitations of such applications. These findings encourage the use of ML to derive chemical insight directly from multidimensional spectroscopy experiments.

\section{Methods}

\subsection{Spectral database for machine learning}

We performed nonlinear response simulations in Python to generate our training and testing datasets. Because of computational costs and storage limitations, we limited the scope of the current study to models for molecular dimers. Studies from the last two decades found that simple molecular models, such as the harmonic oscillator or purely electronic dimer models, are often insufficient to describe sub-picosecond photophysics.\cite{tiwari2013electronic,chenu2013enhancement,tempelaar2014vibrational,fuller2014vibronic} Hence, we used a Holstein-like vibronic exciton Hamiltonian, which was shown to be accurate for predicting features in experimental 2DES spectra of light-harvesting systems.\cite{tempelaar2014vibrational,halpin2014two,schultz2022coupling,schultz2021influence,bakulin2016real,tempelaar2017vibronic} The system Hamiltonian is

\begin{equation}\label{eq:sys_Hamiltonian}
    H_{sys} = H_{el} + H_{vib} + H_{el-vib},
\end{equation}

\noindent where $H_{el}$ and $H_{vib}$ are the electronic and vibrational Hamiltonians, and $H_{el-vib}$ describes the electron-vibrational coupling. The electronic portion of eq \ref{eq:sys_Hamiltonian} for a molecular dimer is written in the Condon approximation as

\begin{equation}\label{eq:el_Hamiltonian}
    H_{el} = \sum_n \epsilon_n c_n^{\dagger}c_n + J_{Coul}\sum_{n\neq n'} c_n^{\dagger}c_{n'} ,
\end{equation}

\noindent where $\epsilon_n$ is the electronic transition energy for molecule $n$, $c_n^{\dagger}$ and $c_n$ are the electronic creation and annihilation operators, respectively, such that $c_n^{\dagger}c_n$ represents an exciton on site $n$, and $J_{Coul}$ is the Coulombic coupling. The vibrational and vibronic Hamiltonians are:

\begin{align}\label{eq:vib_Hamiltonians}
    H_{vib} & = \sum_m \hbar\omega_m b_m^{\dagger}b_m , \\ 
    H_{el-vib} & = \sum_n \sum_m \hbar\omega_m c_n^{\dagger}c_n (\lambda_m (b_m^{\dagger} + b_m) \nonumber \\ 
    & + \lambda_m^2) ,  
\end{align} 

\noindent where $b_{m}^{\dagger}$ ($b_{m}$) creates (annihilates) vibrational quanta for vibration $m$ with frequency $\omega_{m}$ and Huang-Rhys factor $\lambda_{m}^2$. 

\begin{figure}[h]
\centering 
\includegraphics[scale=0.4]{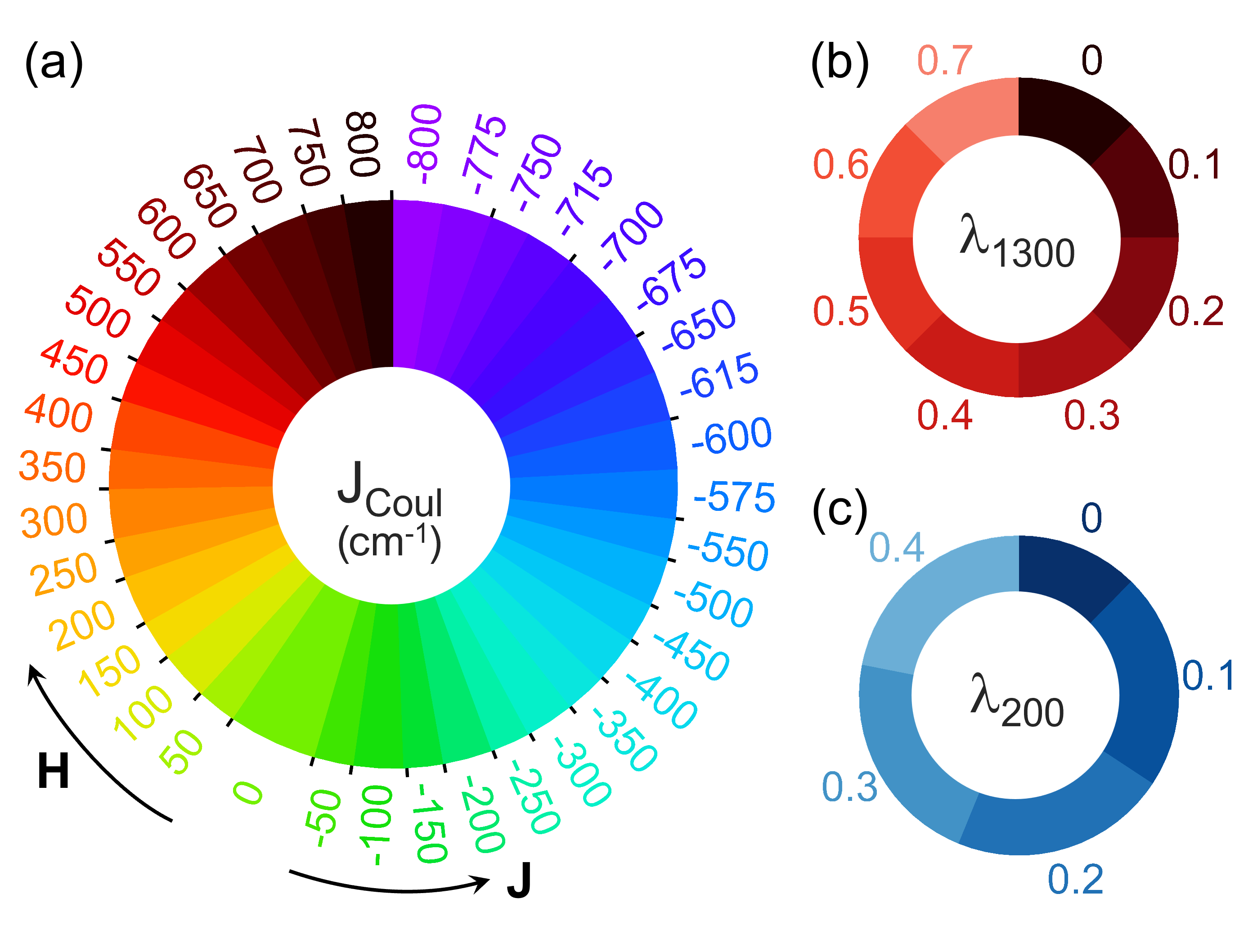}
\caption{Values of the (a) Coulombic coupling ($J_{Coul}$) and nuclear displacements ($\lambda_i$) of the (b) $i = 1300$ and (c) $i = 200$ \wavenumbers \space modes (eqs \ref{eq:el_Hamiltonian} and \ref{eq:vib_Hamiltonians}) used in generating the spectral database. There are 356000 unique 2DES spectra in the full dataset, reflecting 1424 unique homodimers. Slice areas in each hollowed circle are proportional to the amount of data they represent. Outward-facing ticks in (a) indicate the boundaries of the 33 classes reflected in the output of the neural network (\textit{vide infra}). See Table \ref{tab:SI-Ham-parameters} for further details.}
\label{fig:parameter_space}
\end{figure}

In generating the spectral database with the vibronic exciton Hamiltonian (eq \ref{eq:sys_Hamiltonian}), we set ranges for all Hamiltonian parameters so that the simulated spectra correspond to molecular systems that are typically studied with 2DES. Figure \ref{fig:parameter_space} shows the parameter distributions for the Coulombic couplings and the Huang-Rhys factors. We varied the Coulombic coupling from $J_{Coul} = -800$ to $+800$ \wavenumbers \space (Figure \ref{fig:parameter_space}a), which corresponds to strong J- and H-type coupling interactions, respectively.\cite{hestand2018expanded} We previously found that NNs disproportionately misclassify the value of $J_{Coul}$ that underpins the 2DES spectra of J-type dimers.\cite{parker2022mapping,sbaiti2025machine} Thus, while we primarily used a $50$ \wavenumbers \space increment as $J_{Coul}$ was varied, we used smaller increments in $J_{Coul} < 550$ \wavenumbers \space (see Figure \ref{fig:parameter_space}a).

We made two compromises to balance storage costs with the generality of our data space. First, we considered only homodimers (i.e., $\epsilon_1 = \epsilon_2 = \epsilon$ in eq \ref{eq:el_Hamiltonian}). We chose the specific value of $\epsilon = 14500$ \wavenumbers \space to align with the approximate transition energy of terrylenediimide, a prototypical organic chromophore with extensive prior 2DES characterization.\cite{schultz2021influence,schultz2022coupling,zhao2021temperature,mandal2018two,mewes2021broadband} Second, we included two independent vibrations in eq \ref{eq:vib_Hamiltonians}, one high- and one low-frequency (1300 and 200 \wavenumbers, respectively). High-frequency modes, especially C=C stretches, typically exhibit significant Franck-Condon (FC) activity in organic chromophores.\cite{hestand2018expanded} Also, low-frequency modes are found to play significant roles in non-adiabatic excited-state dynamics.\cite{hong2022ultrafast,kim2021low,lin2022accelerating,kim2022pi} Further details of the spectral database are provided in the Supporting Information (SI).

\begin{figure*}[h]
\centering
\includegraphics[scale=0.7]{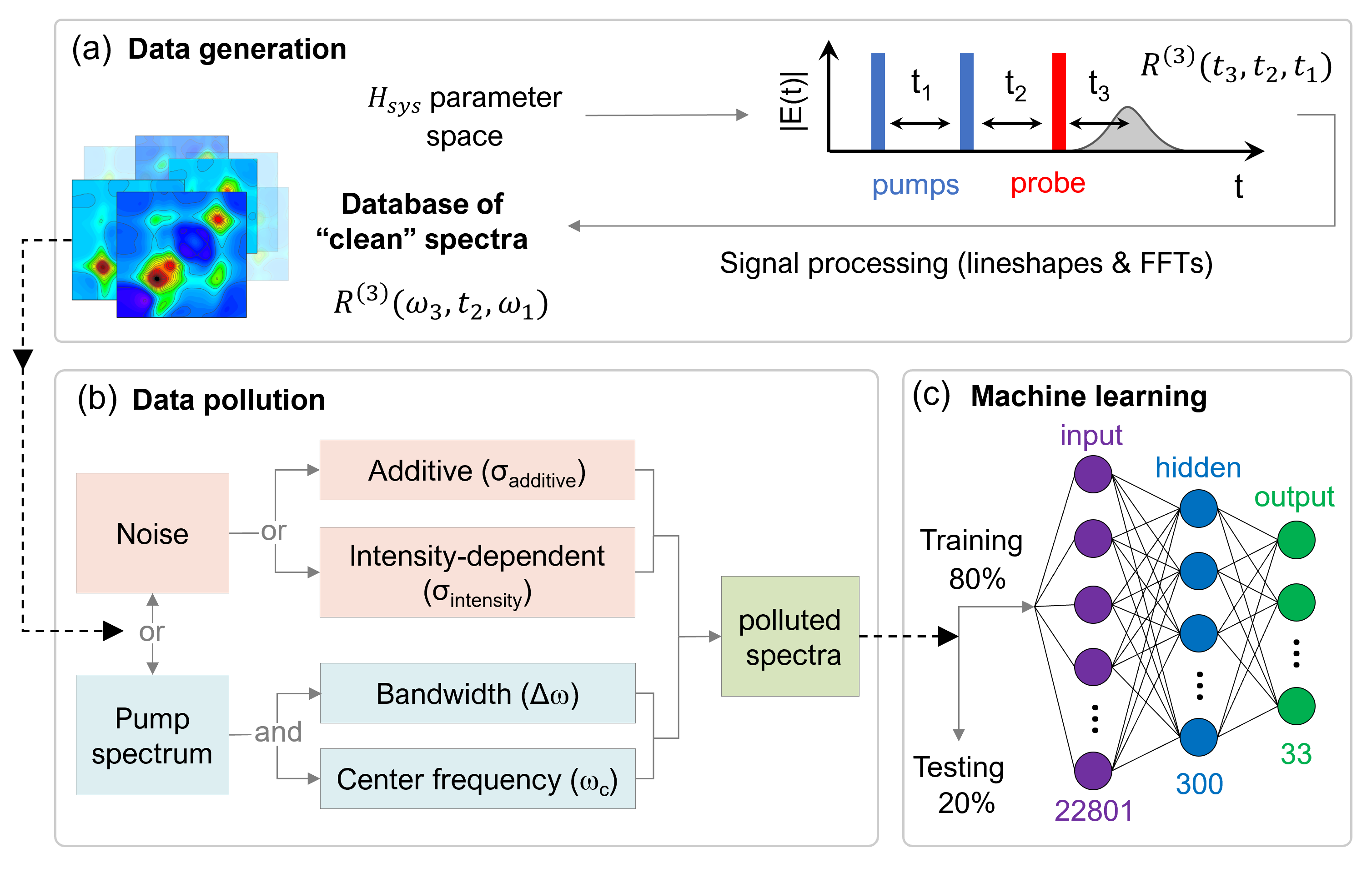}
\caption{Schematic workflow of the spectral simulations, data processing, and machine learning trial employed here. (a) We used nonlinear response function simulations to generate a spectral database for all systems within the parameter space portrayed in Figure \ref{fig:parameter_space}. (b) For each type of data pollutant, we operated on a copy of the clean spectral database and sent the polluted spectra to the ML algorithm. (c) We used 80\% of the data to train a categorical feed-forward neural network and the remaining 20\% for testing.
}
\label{fig:ml_workflow}
\end{figure*}

\subsection{2DES simulations}

\begin{figure}[ht]
\centering 
\includegraphics[scale=0.53]{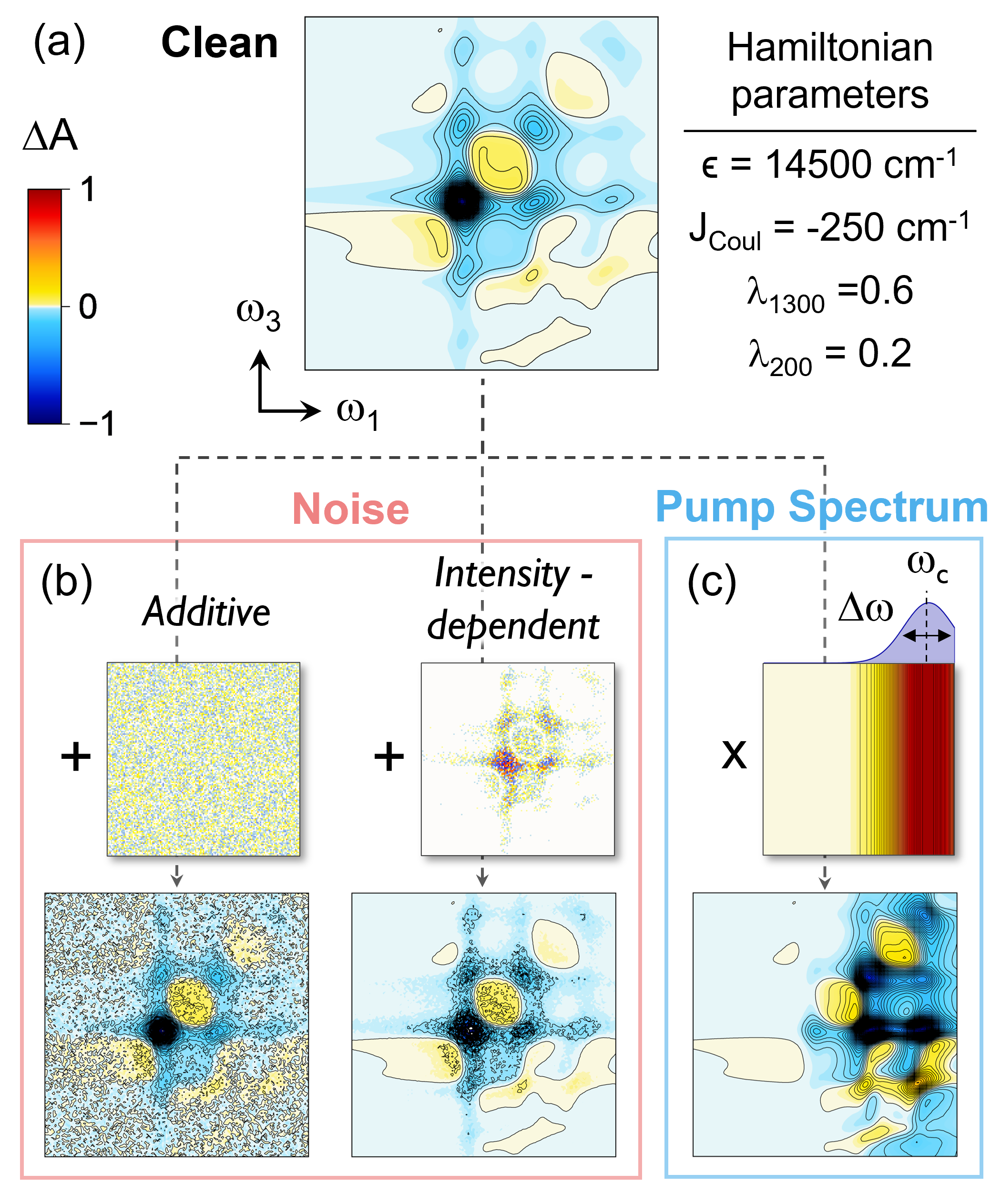}
\caption{(a) A representative ``clean'' spectrum generated with the parameters provided in the inset table. We polluted the datasets by (b) adding one of two types of experimental noise or (c) convoluting the 2DES signal with a Gaussian pump pulse. Representative images of the isolated data pollutants are shown in the upper panels of (b) and (c); the lower panels of (b) and (c) show the resulting polluted spectra. All spectra are plotted against the color scale in (a).}
\label{fig:data-pollution-schematic}
\end{figure}

We simulated absorptive 2DES signals (e.g., Figure \ref{fig:data-pollution-schematic}a) for each model Hamiltonian using in-house Python codes (freely available in Ref. \citenum{schultz2025OREOS}). We calculated the third-order optical response functions (ground-state bleach and stimulated emission pathways) as a function of the $t_1$, $t_2$, and $t_3$ interpulse time delays (Figure \ref{fig:ml_workflow}a). We applied a phenomenological lineshape function\cite{kubo1962resonance} to each dimension of the time-domain signals to account for phenomenological system-bath interactions and to realize finite linewidths. The final absorptive 2DES spectra are computed by fast Fourier transformation of the signal to the pump ($\omega_1 / (2\pi c)$) and probe ($\omega_3 / (2\pi c)$) frequency domains  (abbreviated herein as $\omega_1$ and $\omega_3$, respectively). Table \ref{tab:SI-response-parameters} shows the parameters that were used in our nonlinear response simulations. We selected parameters that reflect common scenarios encountered in 2DES experiments (e.g., spectral linewidths, time and frequency resolutions, etc.). Further details of the simulations are described in the SI.

\subsection{Data pollution}

The simulations described above provide ``clean'' spectra, which do not capture many features of experimentally measured 2DES spectra. Noise and pulse properties can significantly influence the results of 2DES experiments\cite{fuller2015experimental,kearns2017broadband,de2017resolving}, yet such factors are commonly neglected in simulations. To explore: (i) how experimental effects (noise, bandwidth constraints) influence the machine-learnability of 2D data and (ii) bridge simulation-trained NNs toward applications to experimental data, we ``polluted'' our ML datasets prior to both training and testing and examined the resulting effects on NN performance. Figures \ref{fig:ml_workflow}b and \ref{fig:ml_workflow}c show the strategy for introducing each kind (\textit{vide infra}) of pollutant; we applied the pollution operation to a copy of the pristine  dataset, trained the ML on the polluted data, and then computed the performance on a test set of the polluted data (Figure \ref{fig:ml_workflow}c).

Noise signatures, and the spectral characteristics of the pump pulses, are key factors that augment experimental 2DES spectra compared to their simulated counterparts. In 2DES experiments, noise manifests  in many  ways.\cite{saiz2005ensemble} Following established methods to augment spectroscopic data with noise,\cite{saiz2005ensemble} we considered two categories of noise in our study: ``additive noise'' and ``intensity-dependent noise.'' Additive noise represents any source that adds random background noise, such as that introduced by the baseline electronic noise of the detector apparatus.\cite{bressan2023half} While also random, intensity-dependent noise scales with the strength of the 2DES signals. Sources of intensity-dependent noise include shot-to-shot noise from the laser and random fluctuations in the beam overlap at the sample (from fluctuations in temperature, humidity, etc.).

\begin{table}[ht]
\caption{Variables and values therein for each form of data pollutant.}
\centering
\smallskip 
\begin{tabular}{m{1.5cm}|m{1.4cm}|c}
\toprule
\makecell{Data\\Pollutant} & \makecell{Para-\\meter\\(units)} & Values \\
\toprule
\multirow{7}{*}{\makecell{\parbox{1.5cm}{\centering Additive\\ noise}}} & \multirow{7}{*}{\makecell{\parbox{1.4cm}{\centering $\sigma_{additive}$ \\ (unitless)}}} & 0, \num[scientific-notation=true]{0.00001}, \num[scientific-notation=true]{0.000025}\\
& & \num[scientific-notation=true]{0.00005}, \num[scientific-notation=true]{0.000075}\\
& & \num[scientific-notation=true]{0.0001}, \num[scientific-notation=true]{0.00025}\\
& & \num[scientific-notation=true]{0.0005}, \num[scientific-notation=true]{0.00075}\\
& & 0.001, 0.0025, 0.005\\
& & 0.0075, 0.01, 0.025\\
& & 0.05, 0.075, 0.1, 0.25\\
\hline
\multirow{5}{*}{\makecell{\parbox{1.5cm}{\centering Intensity-dependent \\ noise}}} & \multirow{5}{*}{\makecell{\parbox{1.4cm}{\centering $\sigma_{intensity}$ \\ (unitless)}}} & 0, 0.001, 0.0025, 0.005\\
& & 0.0075, 0.01, 0.025\\
& & 0.05, 0.075, 0.1, 0.25\\
& & 0.5, 0.75, 1, 2.5, 5\\
& & 7.5, 10, 25, 50\\
\hline
\multirow{10}{*}{\makecell{Pump \\ spectrum}} & \multirow{3}{*}{\makecell{\parbox{1.4cm}{\centering $\Delta\omega$ \\ (\wavenumbers)}}} & 100, 250, 500, 1000, 1500 \\
                & & 2000, 2500, 3000, 3500 \\
                & & 4000, 5000, 7500, 10000 \\
               \cline{2-3}
               & \multirow{7}{*}{\makecell{\parbox{1.4cm}{\centering $\omega_c$} \\ (\wavenumbers)}} & 12000, 12250, 12500\\
                 & & 12750, 13000, 13250\\
                 & & 13500, 13750, 14000\\
                 & & 14250, 14500, 14750\\
                 & & 15000, 15250, 15500\\
                 & & 15750, 16000, 16250\\
                 & & 16500, 16750, 17000 \\
\bottomrule
\end{tabular}  
\label{tab:imperfection_parameters}
\end{table}

{
\setlength{\parskip}{6pt}

For each unique system Hamiltonian, we modeled noise at every $\omega_1 \times \omega_3 \times t_2$ data point using a normal distribution centered around zero and with a standard deviation of $\sigma$. All 2DES spectra associated with a given model Hamiltonian were normalized to the maximum signal magnitude at $t_2 = 0$. As such, a value of $\sigma = 1$ corresponds to random noise comparable to the signal magnitude (or SNR $\approx 1$ at $t_2 = 0$). Table \ref{tab:imperfection_parameters} provides the values of $\sigma$ that we considered in this study, divided into categories of $\sigma_{additive}$ and $\sigma_{intensity}$. Additive noise is simply added to the 2D spectral data. In contrast, for intensity-dependent noise, we multiply each 2D noise profile (size $n_{\omega_1} \times n_{\omega_3}$, where $n_{\omega_1}$ and $n_{\omega_3}$ are the number of ``pixels'' in the pump and probe frequency dimensions, respectively) element-wise by the 2D spectra prior to addition. See the SI for details of the noise injection procedures.

2DES signals depend critically on the spectral overlap between the pump pulses and the sample absorption. Both the spectral bandwidth ($\Delta\omega$) and center frequency ($\omega_c$) of the pump pulses determine the spectral overlap. To introduce pump pulse characteristics to our ML dataset, we convoluted the simulated 2DES spectra with Gaussian pulses (eq \ref{eq:Gaussian_pulse}) parameterized with realistic values of $\omega_c$ and $\Delta\omega$ (Table \ref{tab:imperfection_parameters}). We defined the former to span the excited-state transition energies of the molecular systems represented in our spectral database (ca. 12000 to 18500 \wavenumbers). Depending on the experimental apparatus, the $\Delta\omega$ of the pump pulses in typical 2DES experiments is typically  1000 - 6000 \wavenumbers.\cite{son2017ultrabroadband,timmer2023full,mewes2021broadband} See the SI for further information.

}

\subsection{Machine learning}

The machine-learning protocols used here are based on earlier workflows of Parker and coworkers\cite{parker2022mapping} that use the PyTorch library\cite{paszke2019pytorch} in Python. Our codes are freely available to the public in Ref. \citenum{schultz2025bridge}. Here, we examined an inverse problem where we trained feed-forward NNs (Figure \ref{fig:ml_workflow}c) to classify 2DES spectra based on the electronic couplings in the underlying model Hamiltonians. The NN uses flattened 2DES spectra (1D arrays of length $n_{\omega_1}\cdot n_{\omega_3}$) as inputs. We used an automated trimming algorithm on all spectra (see the SI for details) to remove outer low-intensity signals and to ensure that all final spectra (i.e. NN inputs) have size: $n_{\omega_1} = n_{\omega_3} = 151$. The NN applies a linear transformation to connect the input layer (consisting of 22,801 neurons from the spectra of size $151 \times 151$ ) with a single hidden layer with 300 neurons. Additional hidden layers produced marginal performance gains, as discussed in the SI. A rectified linear unit (ReLU) activation function is applied to the hidden layer output, followed by a dropout operation for regularization. Finally, a linear transformation and softmax activation function connect the output of the dropout operation to the output layer. Each of the 33 neurons in the output layer corresponds to a single class of electronic coupling $J_{Coul}$ (see Figure \ref{fig:parameter_space}a for the class bounds in the $J_{Coul}$ parameter space). 

\begin{table}[ht]
\caption{Hyperparameters used for all NN trials in this study.}
\centering
    \begin{tabular}{ll}
    Hyperparameter & Value \\
    \midrule
    Activation function & ReLU \\
    Training-testing split & 80:20 \\
    Learning rate\textsuperscript{a} & 0.001 \\
    Number of hidden layers & 1 \\
    Hidden layer size\textsuperscript{a} & 300 \\
    Epochs & 30 \\
    Dropout probability\textsuperscript{a} & 0.2 \\
    Batch size & 100 \\
    \bottomrule
    \end{tabular}
    \begin{minipage}{0.6\linewidth}
        {\footnotesize \raggedright
        \textsuperscript{a}Optimized with a grid search for the unpolluted dataset (see Table \ref{tab:SI-hyperparameter-grid-search}).}
    \end{minipage}
\label{tab:hyperparameters}
\end{table}

We conducted independent ML trials for each polluted dataset (i.e., the NN was trained and tested on each polluted dataset). For simplicity, we determined a set of hyperparameters (Table \ref{tab:hyperparameters}) that optimizes NN performance when trained and tested on clean data. We then kept the hyperparameters constant for all ML trials with the polluted datasets. We also used the same initializations for the trainable parameters (e.g., weights and biases) in each trial and kept the training and testing subsets consistent by seeding data shuffling and splitting operations. See the SI for additional details of our ML procedures. 

In addition to the conventional accuracy metric that assesses NN performance, we used the scikit-learn module\cite{scikit-learn} in Python to calculate F1 scores and top-k accuracies. Compared to accuracy, the F1 score provides better accounting of false positives and false negatives, as well as more robustness to class imbalances.\cite{grandini2020metrics} The top-k accuracy examines whether the true classification is in the top \textit{k} most probable classifications predicted by the NN. Thus, the top-k accuracy provides additional insight into the precision of NN classifications (e.g., how far the misclassifications are from ground truth).

\section{Results and discussion}

The quality of NN classifications when trained and tested on clean (not polluted) spectra is a key reference point for this study. We found that the NN classifies clean 2DES spectra in their correct $J_{Coul}$ category with an accuracy of 83.99\% and a F1 score (macro-averaged) of 0.845. This high performance is consistent with our previous study,\cite{parker2022mapping} in which we found an accuracy of ca. 92\% for a similar $J_{Coul}$ range subdivided into five categories (as opposed to the 33 used here). Note that, in general, we observed that the accuracy and F1 scores were approximately equal (within about one percentage point, as shown in Figure \ref{fig:SI-performance-metrics}). For clarity, we only report the F1 scores.

\begin{figure}[ht]
\centering 
\includegraphics[scale=0.8]{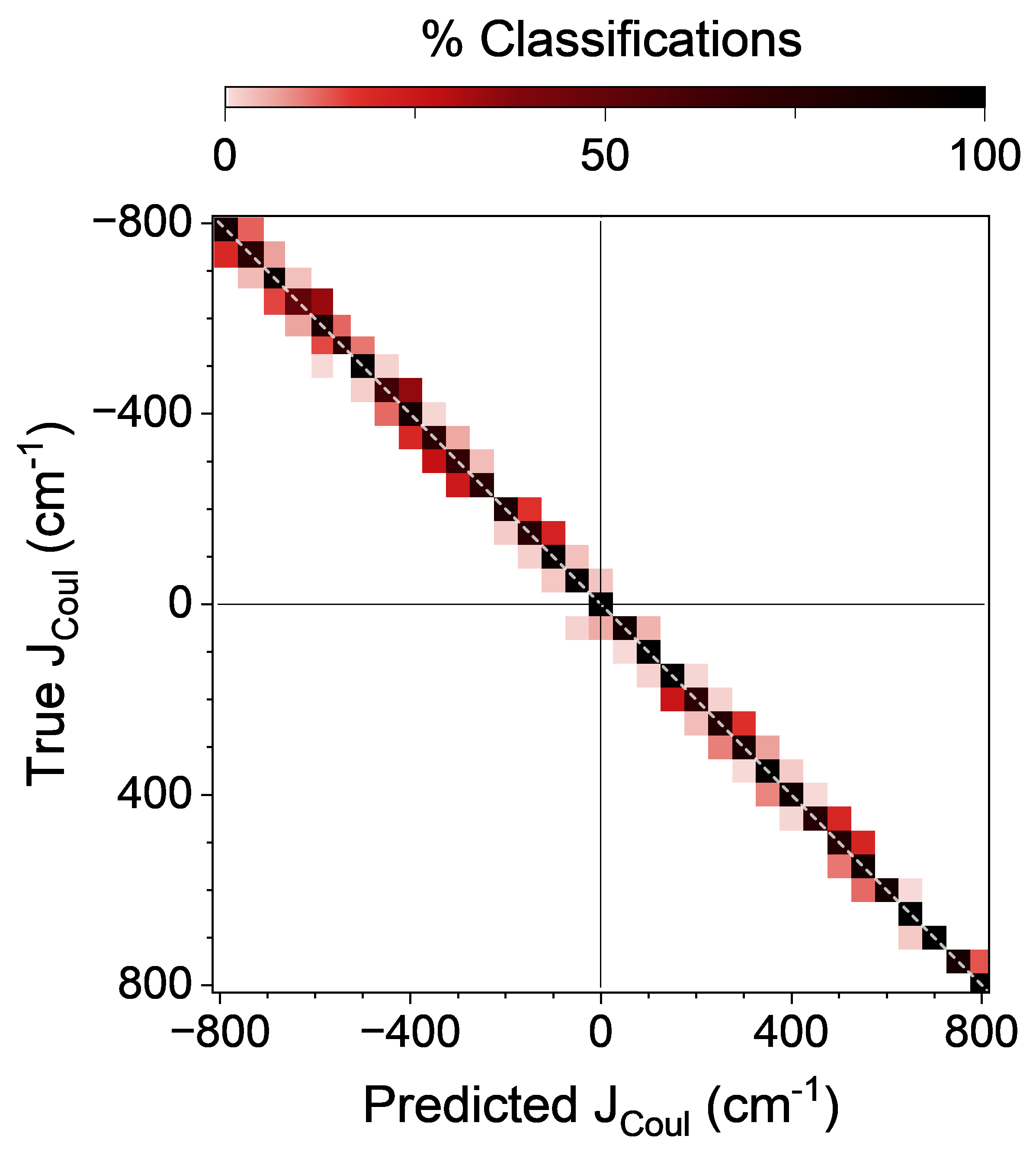}
\caption{Confusion matrix comparing the true vs. NN-predicted values of $J_{Coul}$ when trained and tested on clean data. Each row is normalized to unity. Diagonal entries, indicated by the dotted white line, reflect correct classifications; off-diagonal entries report on misclassifications.}
\label{fig:clean_confusion_mat}
\end{figure}

Figure \ref{fig:clean_confusion_mat} shows the performance of the NN trained and tested using clean spectra through the lens of a confusion matrix. In the confusion matrix representation, correct and incorrect NN classifications are reflected by on- and off-diagonal values, respectively. We observe that while 16\% of the NN classifications are incorrect, the majority of misclassifications occur only one category away from the ground truth. This observation is consistent with the calculated 99.04\% top-2 accuracy.

\subsection{Influence of noise on NN performance}

\begin{figure*}[hb]
\centering 
\includegraphics[scale=0.7]{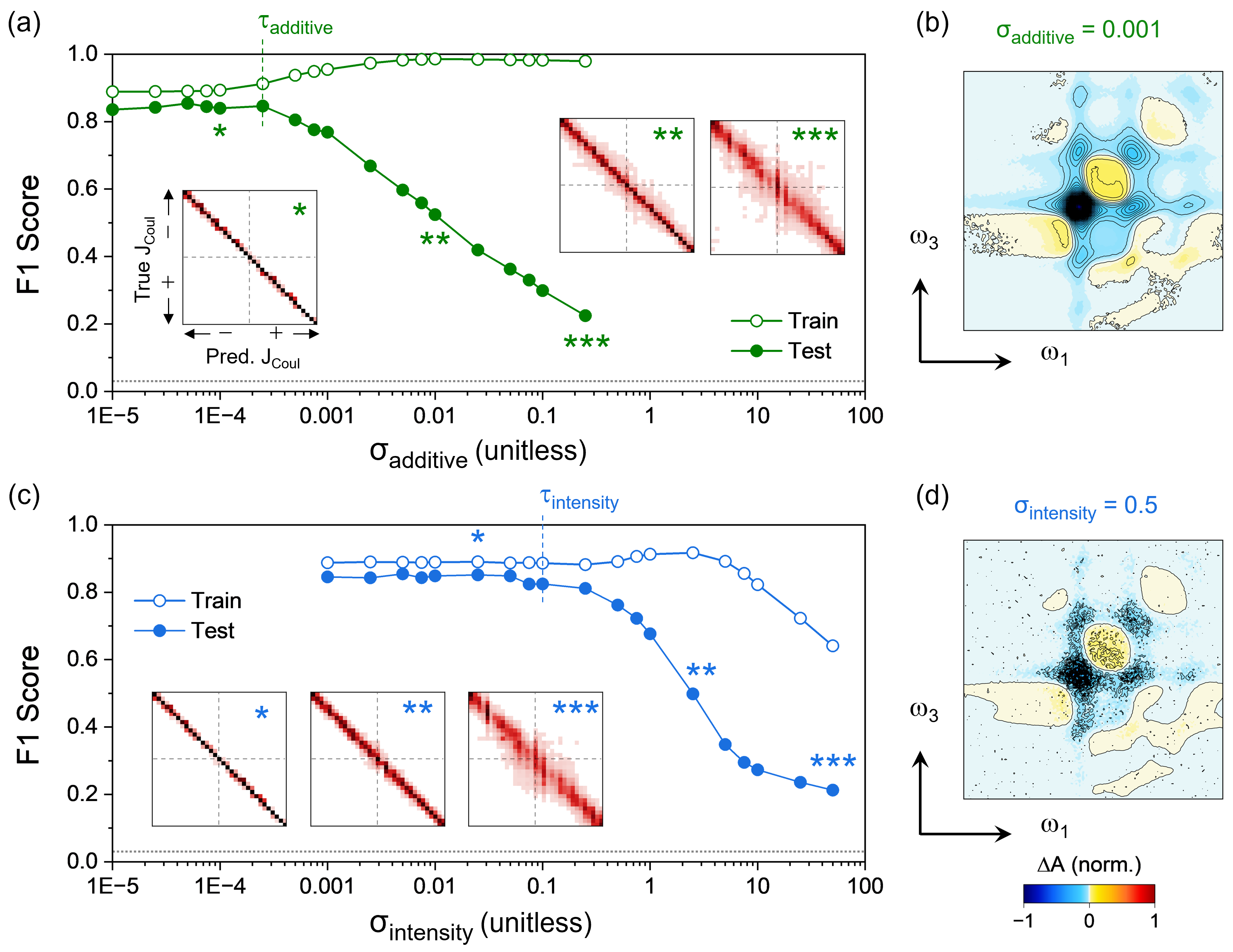}
\caption{Performance of NNs trained and tested on datasets with varying amounts of additive and intensity-dependent noise. (a) and (c) show the F1 scores as a function of $\sigma$ for additive and intensity-dependent noise sources, respectively. (b) and (d) show example 2DES spectra from noisy datasets with $\sigma$ slightly greater than $\tau$. Insets in (a) and (c) show confusion matrices for each of the scenarios denoted by asterisks in the corresponding panels. The confusion matrices are plotted with the same scales as in Figure \ref{fig:clean_confusion_mat}.}
\label{fig:confusion_mats_noise_results}
\end{figure*}

The dependence of the NN performance on the amount of additive noise in the dataset is shown in Figure \ref{fig:confusion_mats_noise_results}a. We find that training and testing F1 scores are unaffected by additive noise until $\sigma_{additive}$ exceeds a threshold of $\tau_{additive} \approx \num{7.5e-4}$ (corresponding to SNR $\approx 6.6$). For example, the testing F1 score drops to 0.779 when $\sigma_{additive} =$ \num{1e-3} (see Figure \ref{fig:confusion_mats_noise_results}b for a representative 2DES spectrum). Above $\tau_{additive}$, the testing F1 score appears to drop exponentially with increasing $\sigma_{additive}$. The confusion matrices inset in Figure \ref{fig:confusion_mats_noise_results}a show that, as $\sigma_{additive}$ increases, NN misclassifications that are more than one category away from the ground truth become increasingly common. The density of off-diagonal elements near the center of the confusion matrix for $\sigma_{additive} =$ \num{0.25} suggests that the NN struggles the most with dimers that have weak-to-intermediate electronic coupling ($-500 \lessapprox J_{Coul} \lessapprox 500$ \wavenumbers) when additive noise is high.

Figure \ref{fig:confusion_mats_noise_results}a also shows that as the amount of additive noise increases, there is an increasingly large gap between the training and testing F1 scores, which is a hallmark signature of over-fitting.\cite{noh2017regularizing} For $\sigma_{additive} \leq \tau_{additive}$, the F1 score of the model on training data is consistently ca. 0.05 above the F1 score on the testing data. This likely indicates a small amount of over-fitting to training data. However, as $\sigma_{additive}$ is increased above the threshold of $\approx \num{7.5e-4}$, the training F1 score rises and plateaus near $0.98$ while the testing F1 quickly drops, indicating over-fitting as the NN `memorizes' the training dataset.

Noise injection is commonly performed\cite{grandvalet1997noise, akbiyik2023data,holmstrom1992using,yin2015noisy} in ML approaches applied to other types of datasets to improve the generalizability of NNs (e.g., to mitigate over-fitting\cite{noh2017regularizing}). However, previous studies\cite{han2020survey,xiao2015learning, bae2024stochastic} found that deep neural networks (DNNs) tend to over-fit when trained on data with noisy labels. This tendency was shown to evince a shift in the DNN from learning general features of the training data to memorizing the noise patterns.\cite{bae2024stochastic} While the trends in Figures \ref{fig:confusion_mats_noise_results}a and \ref{fig:SI-training-performance-additive} suggest a similar effect when the feed-forward NN is trained on spectra with additive noise, note that even for $\sigma = 0.25$, the NN still significantly outperforms random guessing.

As with the ML trials with additive noise, we find that the training and testing F1 scores are unaffected by intensity-dependent noise until a threshold is exceeded (Figure \ref{fig:confusion_mats_noise_results}c), i.e. $ \sigma_{intensity} > \tau_{intensity} \approx  0.5$ (corresponding to SNR $ < 2.5$). The intensity-dependent threshold, $\tau_{intensity}$, is significantly higher than for additive threshold, $\tau_{additive}$. This makes sense, as increasing $\sigma_{additive}$ leads to an increase in SNR more quickly than increasing $\sigma_{intensity}$ (see Figure \ref{fig:SI-SNR-comparison}). Figure \ref{fig:confusion_mats_noise_results}c shows that for $\sigma_{intensity} > \tau_{intensity}$, the NN performance exhibits a logistic-like decay with increasing intensity-dependent noise (in contrast to the exponential decay found for additive noise). In contrast, the training F1 score shows a slight growth from $0.8821$ to $0.9128$ between $\sigma = 0.5$ and $5$, followed by an exponential decay for $\sigma > 0.5$. Aside from evincing over-fitting, this result suggests differences between the nature of over-fitting for spectral datasets with additive vs. intensity-dependent noise.

The results in Figure \ref{fig:confusion_mats_noise_results} together show that NN-based analyses of 2DES data are robust to both additive and intensity-dependent noise sources up to certain thresholds (\num{5e-4} and \num{0.5}, respectively). Above these thresholds, the NNs exhibit a mixture of learning and memorizing, leading to more misclassifications, and many are more than one category away from the true class. This observation is especially true for spectra from Hamiltonians that have weak-to-intermediate electronic coupling values. The threshold values suggest that additive noise sources may be more problematic for NN applications in experimental 2DES contexts compared to intensity-dependent sources. Measurements with $\sigma > \num{1}$ from an intensity-dependent noise source are uncommon in practice, as this scenario suggests fluctuations that commonly exceed the signal magnitude. In contrast, values of $\sigma$ from additive sources in 2DES experiments often exceed \num{5e-4}. In cases like this, common methods for increasing the SNR (e.g., averaging, phase-cycling, etc.) may be necessary to enable the use of NN-based tools to solve inverse problems with the measured spectra.

\subsection{Influence of pump characteristics on NN performance}

\begin{figure*}[htb]
\centering 
\includegraphics[scale=0.63]{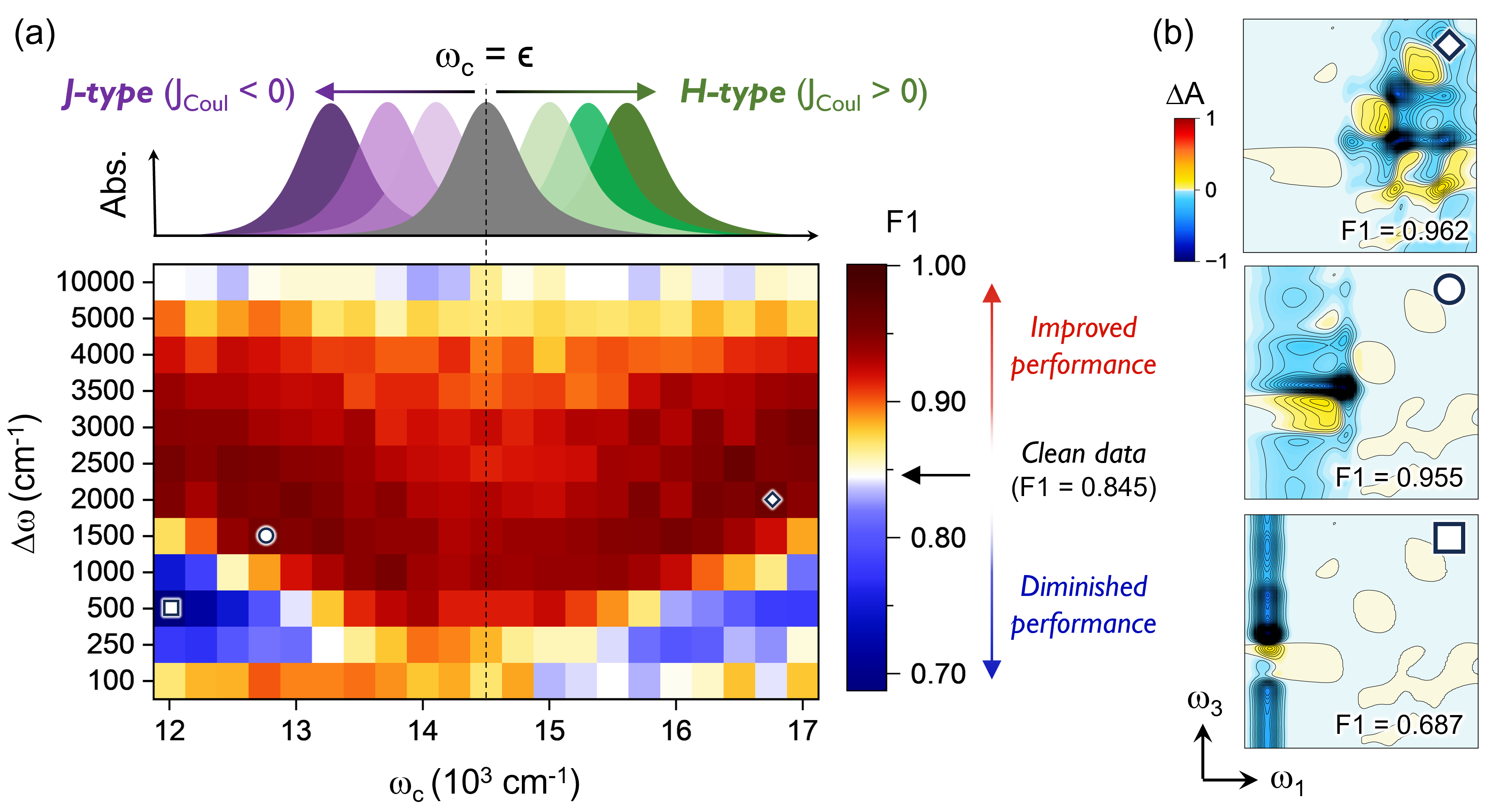}
\caption{(a) NN F1 score for the testing data as a function of $\Delta\omega$ and $\omega_c$ of the pump pulses. The color scale is relative to the F1 score of $0.8448$ found from the clean dataset (red is a higher and blue is lower F1). The upper panel illustrates the expected optical responses of purely electronic J- versus H-type aggregates in Kasha's exciton model. The dashed line indicates exact resonance between the center-frequency of the pump pulses and the monomer optical response. (b) Example 2DES spectra and F1 scores from the corresponding datasets for the ($\Delta\omega$, $\omega_c$) coordinates of the matching shape in (a).}
\label{fig:pump_spectrum_result}
\end{figure*}

Resonance between the pump pulses and the absorption spectrum of the sample in a 2DES experiment critically determines the magnitude and shape of features in the 2DES spectra. As described above and shown in Figure \ref{fig:data-pollution-schematic}c, we varied the spectral bandwidth ($\Delta\omega$) and center frequency ( $\omega_c$) of the pump pulses to simulate experiments with varied resonance conditions. Figure \ref{fig:pump_spectrum_result}a shows the testing F1 score after training and testing our NN with datasets for each $\Delta\omega$ and $\omega_c$ combination.

The heatmap in Figure \ref{fig:pump_spectrum_result}a shows rich variation in the NN performance on the testing data as $\Delta\omega$ and $\omega_c$ of the pump pulses are varied. For all $\omega_c$, the F1 scores when $\Delta \omega = 10000$ \wavenumbers \space are similar to those obtained from the clean dataset (ca. 0.8448). As $\Delta\omega$ decreases, we observe that the F1 scores increase and subsequently decrease. The values of $\Delta\omega$ that yield the maximum F1 score depend strongly on $\omega_c$. Several combinations of $\Delta\omega$ and $\omega_c$ yield F1 scores above 0.95 (dark red regions). Within the range $500 \leq \Delta\omega \leq 5000$ \wavenumbers, the F1 scores are bi-modal with respect to $\omega_c$. For $\omega_c \leq 14000$ \wavenumbers \space and $\geq 15000$ \wavenumbers, small values of $\Delta\omega$ result in F1 scores below the 0.8448 score obtained with the clean dataset. All trends noted in Figure \ref{fig:pump_spectrum_result}a are also found in the training F1 scores (Figure \ref{fig:SI-dual-pump-training}).

The dependencies of the F1 score on $\Delta\omega$ and $\omega_c$ are counter-intuitive for two reasons. First, 2DES experiments are typically designed with maximal pulse bandwidth.\cite{son2017ultrabroadband,kearns2017broadband} This is because lower values of $\Delta\omega$ constrain the shape of the 2DES signal along the pump axis, in turn obscuring information about the molecular system. For example, compare the upper and middle spectra in Figure \ref{fig:pump_spectrum_result}b to Figure \ref{fig:data-pollution-schematic}a. In contrast, we find that smaller $\Delta\omega$ values improve NN performance (to a limit). Also, we might expect better NN performance when the pump pulse spectra are resonant with the diabatic excited-state energy of the monomers in eq \ref{eq:el_Hamiltonian} (i.e., $\omega_c = \epsilon = 14500$ \wavenumbers). Instead, we find that, for almost all $\Delta\omega$ values, the F1 scores increase when the pump spectra are significantly red- or blue-shifted away from the monomer transition energy.

Kasha’s theory\cite{kasha1963energy,kasha1965exciton,hestand2018expanded} for the optical responses of molecular aggregates predicts two exciton classifications based on the sign of the Coulombic coupling. The theory predicts that the absorption spectrum of a dimer with $J_{Coul} < 0$ (J-type) will be red-shifted compared to that of the isolated monomer (illustrated in the upper portion of Figure \ref{fig:pump_spectrum_result}a). In contrast, dimers with $J_{Coul} > 0$ (H-type) yield blue-shifted absorption spectra. The qualitative predictions of Kasha's theory correlate well with  both  (i) the bimodal dependence of NN accuracy on $\omega_c$ and (ii) the symmetry of the bimodal trend about $\omega_c = \epsilon = 14500$ \wavenumbers. Such a correlation makes sense since, for all pump spectra except those with $\omega_c = 14500$ \wavenumbers, the pump biases the spectral dataset toward one exciton response regime and, in turn, influences how the NN learns about the underlying electronic couplings. From the trends in the F1 score as $\Delta\omega$ is varied, we posit that, for sufficiently large $\Delta\omega$, biasing one exciton regime over the other boosts NN performance by emphasizing the differences in the 2DES signatures of H- vs. J-type aggregates. However, as $\Delta\omega$ is decreased, the performance gains from biasing one exciton regime should eventually be overcome by the erasure of information contained in off-resonant regions of the 2DES spectra. We observe this behavior for all $\omega_c$, as the NN performance drops substantially when $\Delta\omega$ drops below threshold values (e.g., $\tau_{\Delta\omega} = 1500$ \wavenumbers \space for $\omega_c = 12250$ \wavenumbers).

The findings of Figure \ref{fig:pump_spectrum_result} show that feed-forward NNs more accurately map 2DES spectra to electronic couplings when the datasets are spectrally constrained (polluted by pump resonance). This result marks a significant departure from human-based designs and analyses of 2DES experiments. With few exceptions,\cite{senlik2015two} spectrally broadband and on-resonance pump pulses are desired for 2DES experiments. Heisler and coworkers\cite{de2017resolving} showed that limited resonance between the pump pulses in a 2DES experiment and the absorption spectrum of the molecular monomer can artificially manifest signatures of electronic coherences in the spectra, which are physically impossible for monomeric samples. While such unphysical information may mislead human analysis of 2DES data, our findings show that constraining spectral resonance positively influences the ability of NNs to learn about spectral signatures of electronic coupling.

\subsection{Implications for applications to 2DES experiments}

ML presents revolutionary opportunities for decoding information from optical data.\cite{fang2021decoding,meza2021applications} The results of our study find that, despite the signal complexity of nonlinear multidimensional spectroscopy, simple ML approaches like the feed-forward NNs can learn information about the underlying molecular properties in the face of experimental realities (noise and pulse resonance conditions). Although additive and intensity-dependent noise signatures both degrade NN performance and lead to overfitting, this is only the case for noise widths that exceed some threshold ($\sigma > \tau$). Here, we found that the NN remains unhampered until the SNR drops below $6.6$ for additive noise and $2.5$ for intensity-dependent noise. This result implies that (i) sources of intensity-dependent noise pose limited risk of obscuring coupling information in experimental 2DES spectra, and (ii) that using NNs to map couplings from experimental data should be robust to noise if the SNR is sufficiently high. Methods to mitigate the negative effects of noise on the machine learnability of experimental 2DES data, such as averaging or phase cycling,\cite{zhang2012phase} may be necessary for ML-led inverse problem solving with excessively noisy data.

The counterintuitive behavior revealed in Figure \ref{fig:pump_spectrum_result} highlights that NNs interface with spectroscopic signals in a fundamentally different way compared to humans. We hypothesize that ML tools may provide opportunities to leverage subtle properties of the multidimensional spectra that are overlooked by traditional interpretation methods. For example, the traditional workflow to interpret 2DES spectra for complex molecular systems follows insights gained from nonlinear optical response theories.\cite{mukamel1995principles,cho2008coherent,biswas2022coherent} Theoretical models predict that cross-peaks in rephasing 2DES spectra are particularly sensitive to electronic and vibronic couplings.\cite{perlik2014distinguishing,tiwari2013electronic,halpin2014two,dean2016vibronic} In turn, cross-peaks are of central focus in the analysis of experimental 2DES data.\cite{wang2017controlling,dean2016vibronic,thyrhaug2018identification,policht2022hidden} The salient trends in the predictions from nonlinear optical response theories tend to guide human-based analyses of spectra, but there may be a wealth of information contained in the fleeting trends in theoretical predictions. Our observation that the NN-interpretability of 2DES data is maximized by sub-optimal (by human standards) resonance conditions supports our hypothesis. 

NNs elicit an \textit{information-centric} perspective of spectroscopic signals during training. In a recent study of Flores and coworkers,\cite{rieger2023understanding} the authors trained a CNN to classify linear infrared spectra based on functional group information. In addition to spectral features from fundamental vibrational frequencies, they found that the model uses non-intuitive features, such as the absence of specific peaks or peaks from anharmonic modes, in its classifications. Such findings emphasize the potential usefulness of traditionally overlooked properties of spectra in enabling accurate spectral interpretations. Our findings prompt further explorations of how property-specific information is distributed throughout 2DES datasets. Indeed, a recent study of Jakobsson and coworkers\cite{bolzonello2024fisher} found patterns of Fisher information distribution in simulated 2DES spectra that differ from the typical spectral regions that nonlinear response theories suggest for analysis.\cite{perlik2014distinguishing,tiwari2013electronic,halpin2014two,dean2016vibronic} Information-based (machine-learned in our case) approaches may guide experimental designs or spectral analyses that most efficiently lead to molecular insight from multidimensional spectra.

\section{Conclusion}

2DES spectroscopy is an increasingly accessible and powerful tool that can probe ultrafast dynamics. Chemically meaningful information is traditionally inferred from 2DES spectra through extensive signal analysis,  theoretical modeling, and human-led comparisons of simulated and experimental spectra.\cite{halpin2014two,thyrhaug2018identification,policht2022hidden} Despite the time and effort required to perform such tasks, misinterpretations of 2DES spectra are still possible and are historically precedented.\cite{cao2020quantum,duan2017nature,manvcal2020decade,zerah2021photosynthetic,schultz2024coherence} Misinterpretations pose a concern, especially as 2DES is used to study  increasingly complicated condensed-phase systems. Being agnostic to traditional strategies for interpreting spectra, ML offers a promising route to translate experimental spectra to chemical insight in a robust and data-driven manner. Indeed, there are few studies\cite{namuduri2020machine} that use ML as an inverse problem solving tool to address experimental 2DES data. 

We have showed that even when practical limitations such as noise and pulse resonance conditions are included in the spectral data, feed-forward NNs match simulated 2DES spectra to electronic coupling strengths with high accuracy. We found that additive (e.g., dark noise) and intensity-dependent (e.g., laser power fluctuations) noise signatures degrade NN performance after threshold amounts of noise are exceeded. The threshold for intensity-dependent noise is significantly higher than for additive noise, suggesting that the former poses a smaller risk of obscuring information about electronic couplings in experimental 2DES spectra. Both kinds of noise lead to over-fitting, which aligns with findings of earlier studies of noise with deep neural networks.\cite{yin2015noisy,xiao2015learning,bae2024stochastic,han2020survey} Our results suggest that methods to mitigate the negative effects of additive noise on the machine learnability of experimental 2DES data, such as averaging or phase cycling, may be necessary to map spectra to molecular properties.

The results presented here convey positive prospects for adapting ML-based tools to analyze and interpret complex experimental 2DES data. Future directions toward ML-guided analyses of experimental spectra may combine polluted simulated data with established transfer learning techniques.\cite{han2021transfer,han2024ai,ren2022machine,wu2024unraveling,ye2025ai,alberts2024leveraging} A potential approach could start with pretraining on polluted simulated spectra to produce a general ML model. Other research groups could then perform retraining (also called fine-tuning)\cite{han2021transfer,han2024ai,alberts2024leveraging} on the final layers of the general model with local, smaller experimental datasets. Transfer learning techniques have shown promising results in other multidimensional spectroscopy studies focused on protein structure classification.\cite{ren2022machine,wu2024unraveling,ye2025ai}

Finally, this study reveals significant differences between the human- and machine-based interpretation of 2DES signals. In contrast to human-based analysis, we found that NNs exhibit enhanced performance (exceeding an F1 score of 0.96) when the data are constrained by the bandwidth and center-frequency of the pump. We attribute such counterintuitive behavior to the pulse resonance \textit{changing} how the NN learns the optical properties of molecular excitons. In other words, biasing the spectral data in either of the exciton absorption regimes (J- or H-type) helps the NN learn how couplings manifest in the spectra. This observation provides evidence that NNs accrue a radically different, more information-centric perspective of electronic coupling signatures in 2DES spectra. Further studies of the machine learnability of CMDS spectra may afford guidelines for experimental design as well as approaches to interpret experimental datasets.

\vspace{3mm}

\begin{center}
{\large \textbf{Acknowledgments}}
\end{center}
J.D.S. gratefully acknowledges support from an Arnold O. Beckman Postdoctoral Fellowship in the Chemical Sciences (dx.doi.org/10.13039/100000997).  Support to K.A.P., B.S., and D.N.B. from the Department of Energy (DE-SC0019400) is acknowledged gratefully.

\vspace{3mm}

\begin{center}
{\large \textbf{Data and code availability}}
\end{center}
The codes used for this work are freely available at two public repositories. The machine learning code, iterative data pollution workflow, and a subset of the training/testing dataset are available at doi.org/10.5281/zenodo.15041004 (Ref. \citenum{schultz2025bridge}). The spectral simulation code used to generate the machine learning dataset is available at doi.org/10.5281/zenodo.6757663 (Ref. \citenum{schultz2025OREOS}).

\vspace{3mm}

\begin{center}
{\large \textbf{Competing interests}}
\end{center}
The authors declare no competing interests.

\clearpage

\onecolumn  
\begin{center}
{\Large \textbf{Supporting Information}}
\end{center}
\appendix

\renewcommand{\thesection}{S\arabic{section}} 
\renewcommand{\thesubsection}{S\arabic{section}.\arabic{subsection}}
\renewcommand{\thefigure}{S\arabic{figure}}
\renewcommand{\thetable}{S\arabic{table}}
\renewcommand{\theequation}{S\arabic{equation}}

\setcounter{equation}{0}  
\setcounter{table}{0}  
\setcounter{figure}{0}  

\section{Additional simulation details}
\label{sec:SI-Additional-simulation-details}

\subsection{Parametrization of the vibronic dimer Hamiltonian}

Table \ref{tab:SI-Ham-parameters} details the parameters we used to construct a set of 1424 unique vibronic dimer Hamiltonians. We formulate the Hilbert space for each system with kets of the form

\begin{equation} 
\label{eq:kets}
    \ket{n_{ea}, n_{v_{1}a}, n_{v_{2}a}, n_{eb}, n_{v_{1}b}, n_{v_{2}b}}
\end{equation}

\noindent where $n_{ei}$ is the electronic quantum number for molecule $i$ ($i = a,b$) and $n_{v_{k}i}$ is the vibrational quantum number for vibrational mode $k$ ($k = 1, 2$ for the $1300$ and $200$ \wavenumbers \space modes, respectively) for molecule $i$. For each independent vibrational mode, we constrained the maximum vibrational quanta in each ket to five.

\renewcommand{\arraystretch}{1.5}
\begin{table}[h]
\caption{All parameters used to yield the 1424 unique vibronic dimer Hamiltonians. Figure \ref{fig:parameter_space} of the main text provides a complementary graphical representation.}
\centering
    \begin{tabular}{C{2.5cm}|C{1.5cm}|C{8cm}}
    \toprule
    \centering Parameter & Units & Values ($N_{systems}$) \\
    \hline
    $\epsilon$ & \wavenumbers & 14500 (1424) \\
    \hline
    $J_{Coul}$ & \wavenumbers & 
    -800 (40), -775 (40), -750 (32), -715 (40), -700 (40), -675 (40), -650 (32), -615 (40), -600 (40), -575 (40), -550 (32), -500 (40), -450 (32), -400 (40), -350 (32), -300 (40), -250 (32), -200 (40), -150 (32), -100 (40), -50 (32), 0 (72), 50 (32), 100 (40), 150 (32), 200 (40), 250 (32), 300 (40), 350 (32), 400 (40), 450 (32), 500 (40), 550 (32), 600 (40), 650 (32), 700 (40), 750 (32), 800 (40) \\
    \hline
    $\lambda_{1300}$ & unitless & 0.0 (178), 0.1 (178), 0.2 (178), 0.3 (178), 0.4 (178), 0.5 (178), 0.6 (178), 0.7 (178) \\
    \hline
    $\lambda_{200}$ & unitless & 0.0 (176), 0.1 (312), 0.2 (312), 0.3 (312), 0.4 (312) \\
    \bottomrule
    \end{tabular}  
\label{tab:SI-Ham-parameters}
\end{table}

\subsection{Simulations of multidimensional spectra}

We used the nonlinear response function formalism, in which the nonlinear molecular response function is calculated from a combination of different pathways in Liouville space.\cite{mukamel1995principles} The transition dipole operator for a light-matter interaction is written\cite{halpin2014two} in the Condon approximation as,

\begin{equation} \label{eq:dipole_operator}
    \mu (\tau_i) = c^\dagger + c
\end{equation}

\noindent where $\tau_i$ reflects the instantaneous time at which the impulsive light-matter interaction occurs.\cite{halpin2014two,jansen2006nonadiabatic} Free propagation of the wavefunction under the system Hamiltonian during the time between two light-matter interactions $j$ and $k$ is achieved with the time-evolution operator,

\begin{equation} 
\label{eq:time-evolution_operator}
    U (\Delta t_{jk}) = e^{-i H_{sys} \Delta t_{jk}} .
\end{equation}

To lower the computational cost of time-propagation, we partition the transition dipole and time-evolution operators into blocks based on the matrix indices of the electronic manifolds. For a generic operator $O$, the operator $O_{jk}$ is a subspace of $O$ corresponding to the $j$ and $k$ block indices. This notation ensures that,

\begin{align}
    \label{eq:mu-example}
    \mu_{ge} \ket{g(\tau_j)} &= \ket{e(\tau_j)} \\
    \label{eq:U-example}
    U_{ee}(\Delta t_{jk}) \ket{e(\tau_j)} &= \ket{e(\tau_k)}
\end{align}

\noindent where we have represented the $S_0$ and $S_1$ states with \textit{g} and \textit{e}, respectively. \\

Forcing $\tau_j$ and $\tau_k$ to represent sequential moments in time ($j = k+1$), we simplify the notation with $\Delta t_{jk} = t_k$, in turn recovering the common notation of the coherence ($t_1$), waiting ($t_2$), and rephasing ($t_3$) time delays in 2DES. The third-order nonlinear response functions are thus,\cite{jansen2006nonadiabatic}

\begin{align}
    \label{eq:R1}
    R_1(t_1,t_2,t_3) &= \bra{i} \UGGdag{t_1} \UGGdag{t_2} \UGGdag{t_3} \muEGdag \UEE{t_3} \muGE \UGG{t_2} \muEGdag \UEE{t_1} \muGE \ket{i} \\[10pt]
    \label{eq:R2}
    R_2(t_1,t_2,t_3) &= \bra{i} \muGEdag \UEEdag{t_1} \UEEdag{t_2} \muEG \UGGdag{t_3} \muEGdag \UEE{t_3} \UEE{t_2} \muGE \UGG{t_1} \ket{i} \\[10pt]
    \label{eq:R3}
    R_3(t_1,t_2,t_3) &= \bra{i} \muGE \UEEdag{t_1} \muEG \UGGdag{t_2} \UGGdag{t_3} \muEGdag \UEE{t_3} \muGE \UGG{t_2} \UGG{t_1} \ket{i} \\[10pt]
    \label{eq:R4}
    R_4(t_1,t_2,t_3) &= \bra{i} \UGGdag{t_1} \muGEdag \UEEdag{t_2} \muEG \UGGdag{t_3} \muEGdag \UEE{t_3} \UEE{t_2} \UEE{t_1} \muGE \ket{i}
\end{align}

\noindent where $R_n$ is the response function for Liouville pathway $n$, and $\ket{i}$ is the initial state. For simplicity, we assume that the system begins in the global ground state (i.e., $\ket{i} = \ket{0, 0, 0, 0, 0, 0}$, following the form of eq \ref{eq:kets}). Eqs \ref{eq:R1} and \ref{eq:R4} correspond to the non-rephasing ground-state bleach (GSB) and stimulated emission (SE) pathways, respectively, while eqs \ref{eq:R2} and \ref{eq:R3} are the rephasing GSB and SE pathways, respectively. \\

We simulated all spectra in the rotating frame\cite{halpin2014two,hamm2011concepts} by removing one or two electronic quanta from the diagonal entries of the blocks corresponding to the singly excited manifold of the electronic Hamiltonian. We included phenomenological effects of system-bath interactions with the lineshape function,

\begin{equation} 
\label{eq:lineshape}
    g(t) = \Delta E^2 t_c^2 e^{-\frac{t_i}{t_c}+(\frac{t_i}{t_c}-1)}
\end{equation}

\noindent where $\Delta E$ captures energy gap fluctuations with correlation time $t_c$.\cite{kubo1962resonance} We incorporated the lineshape function $g(t)$ by multiplying $R_{i=1,2,3,4}(t_1,t_2,t_3)$ along each time dimension with $e^{-g(t)}$. Following fast Fourier transformations along $t_1$ and $t_3$, we calculated absorptive 2DES spectra with,

\begin{equation} \label{eq:Abs}
    R_{Abs}(\omega_1,t_2,\omega_3) = \sum_{i=1}^4 \operatorname{Re}[R_i(\omega_1,t_2,\omega_3)].
\end{equation}

\noindent For all parameters of the simulations, including those of the finite lineshapes, we defined values (Table \ref{tab:SI-response-parameters}) to reflect typical conditions of 2DES experiments (see Refs. \citenum{halpin2014two}, \citenum{schultz2022coupling}, and \citenum{thyrhaug2018identification}, for example).

\renewcommand{\arraystretch}{1}
\begin{table}[h]
\caption{All parameters used to generate 2DES spectra from the 1424 vibronic dimer Hamiltonians.}
\centering
    \begin{tabular}{C{2.5cm}|C{1.5cm}|C{3cm}}
    \toprule
    \centering Parameter & Units & Values \\
    \hline
    $t_1$ & fs & $[0:3:186]^{\text{a}}$ \\
    $t_2$ & fs & $[0:5:1245]^{\text{a}}$ \\
    $t_3$ & fs & $[0:3:186]^{\text{a}}$  \\
    $n_{pad}$ & unitless & $256^{\text{b}}$ \\
    \hline
    \multirow{2}{*}{\makecell{\parbox{2.5cm}{\centering $\Delta E$}}} & \multirow{2}{*}{\makecell{\parbox{1.5cm}{\centering \wavenumbers}}} & 1300 ($t_1,t_3$)\textsuperscript{c} \\
    & & 125 ($t_2$)\\
    \multirow{2}{*}{\makecell{\parbox{2.5cm}{\centering $t_c$}}} & \multirow{2}{*}{\makecell{\parbox{1.5cm}{\centering fs}}} & 40 ($t_1,t_3$)\textsuperscript{c} \\
    & & 300 ($t_2$)\\
    \bottomrule
    \end{tabular}
    \vspace{2mm}
    \begin{minipage}{0.6\linewidth}
        {\footnotesize \raggedright
        \textsuperscript{a} Format: [minimum value: step size: maximum value].
        \newline \textsuperscript{b} Length of zero padding prior to FFT operations.
        \newline \textsuperscript{c} Optical coherences during $t_1$ and $t_3$ dephase faster than coherences during $t_2$. Hence, we scaled the lineshape parameters accordingly.}
    \end{minipage}
    
\label{tab:SI-response-parameters}
\end{table}

\subsection{Automated image resizing}

\begin{figure}[h]
\centering
\includegraphics[scale=0.6]{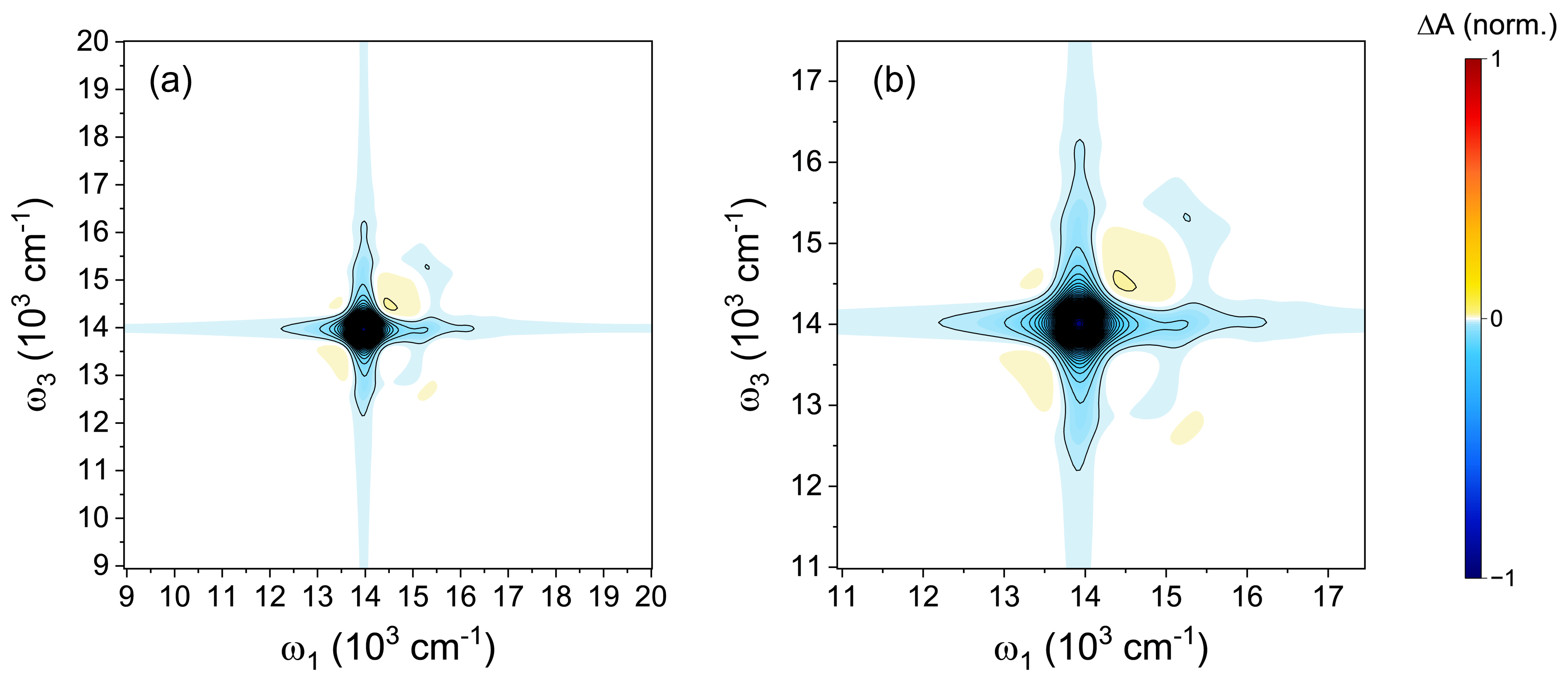}
\caption{Example spectra (a) before and (b) after the automated trimming and centering algorithm. The example spectra correspond to a system Hamiltonian with parameter values: $J_{Coul} = -500$ \wavenumbers, $\lambda_{1300} = 0.2$, $\lambda_{200} = 0$.}
\label{fig:trimming-centering-example}
\end{figure}

The simulations described in the previous section yield spectra of size $256 \times 256$ along the $\omega_1 \times \omega_3$ dimensions (corresponding to approximately 11075 \wavenumbers \space along each frequency axis). Due to the energy scales of the system Hamiltonians and the optical response simulation parameters, there is inherently low signal and therefore low information in the outer regions of the spectra (e.g., Figure \ref{fig:trimming-centering-example}a). To generate smaller, more computationally tractable spectra inputs to the NN, and to remove spectral regions with low information, we used an in-house script to automatically trim the spectra around a central coordinate ($\omega_{1c},  \omega_{3c}$). The algorithm produces ``resized'' spectra with size $151 \times 151$ (approximately $6500$ \wavenumbers \space along each frequency axis; see Figure \ref{fig:trimming-centering-example}b), as mentioned in the main text.

The trimming method works by first determining a central coordinate where the signal is most concentrated ($\omega_{1c},  \omega_{3c}$). The trimmed spectra are then generated by collecting the desired sized subset of data around the central coordinate. It is possible for the new spectra generated by our resizing algorithm to include indices outside the $\omega_1 \times \omega_3$ bounds of the original spectra; this would be the case, for example, if the signals in a given spectrum were highly concentrated in a corner with very low signal everywhere else. In this case, the new pixels included in the spectrum would have zero signal. We did not note any cases where this occurred in our data sets as our spectra did not have this type of signal distribution.

\FloatBarrier

\section{Additional machine learning details}
\label{sec:SI-additional-ML-details}

\subsection{Architecture}

As described in the main text, we extended the ML workflows of Parker and coworkers.\cite{parker2022mapping} Our ML framework relies on the PyTorch library\cite{paszke2019pytorch} in Python. We herein use PyTorch notation to describe the relevant functions, sub-libraries, etc. in our ML methods. We employed the Adam optimizer (torch.optim.Adam) to minimize the CrossEntropyLoss cost function (torch.nn.CrossEntropyLoss). We identified the model predictions as the class index with the highest score (torch.max), and computed probabilities with the Softmax function (torch.nn.Softmax) along the class dimension. Prior to conducing performance analyses (e.g., F1 scores and top-k accuracies), we converted the probabilities output from torch.nn.Softmax() to a NumPy array.

\subsection{Reproducibility}

We used seeds in this work to enable reproducibility between runs. We initialized the trainable parameters of the NN consistently by setting the PyTorch seed (torch.manual\_seed) in all runs to \num{2942}. Prior to splitting the dataset in to training and testing subsets, we shuffled the dataset with NumPy's random number generation (RNG) sub-library, which we seeded with $72067$. Thus, all ML trials used the same subsets of the dataset for training and testing. We also seeded the NumPy RNG for generating Hamiltonian-specific noise profiles while maintaining the Hamiltonian-noise profile correspondence between different ML trials (see Section \ref{subsec:random_noise_injection} for further details).

\subsection{Hyperparameter optimization}

\begin{figure}[h]
\centering
\includegraphics[scale=0.55]{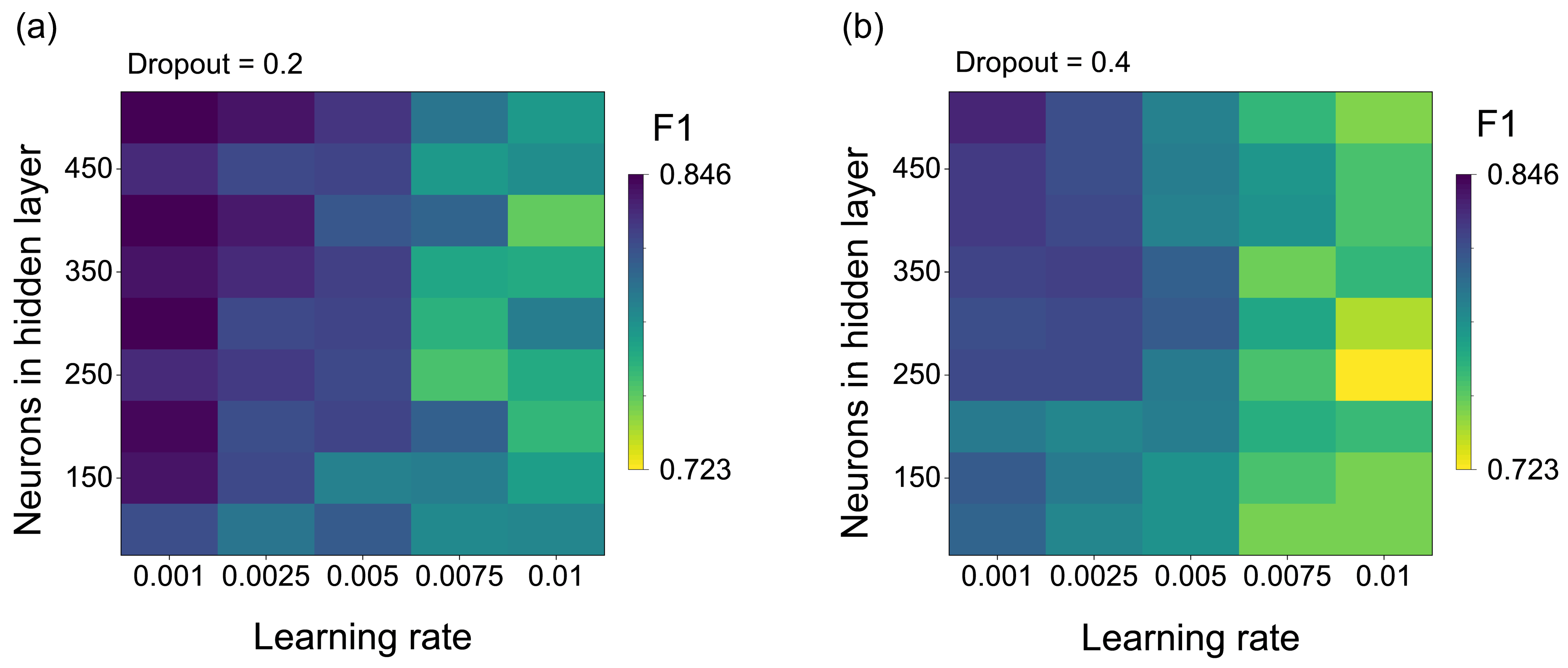}
\caption{Test F1 scores (macro-averaged) as a function of hidden layer size and learning rate with dropout equal to (a) 0.2 and (b) 0.4.}
\label{fig:SI-hyperparameters}
\end{figure}

We performed a grid-search to determine optimal values for the hidden layer size, learning rate, and dropout hyperparameters. Table \ref{tab:SI-hyperparameter-grid-search} provides all the values of the grid search and Figure \ref{fig:SI-hyperparameters} shows model performance for each combination. We found that the parameter set, abbreviated as [hidden layer size, learning rate, dropout], of [500, 0.001, 0.2] yielded the maximum F1 score. This set differs from the parameter set [300, 0.001, 0.2] that we used for this study due to our choice to trade a minor performance loss for a smaller hidden layer (F1 $= 0.8448$ and $0.8457$ for 300 and 500 neurons, respectively), in turn enabling faster training times. For the number of epochs, we examined the behavior of the loss function versus epoch and determined 30 epochs to be a sufficient compromise between high testing accuracy and tractable training times (Figure \ref{fig:SI-training-loss}). To avoid the massive computational cost required to optimize hyperparameters for each uniquely polluted dataset, we performed the grid search solely on the clean dataset and kept the resulting hyperparameters constant for all other ML trials.

\begin{table}[h]
\caption{Parameters of hyperparameter grid search}
\setlength\tabcolsep{6pt}
\centering
    \begin{tabular}{C{2.5cm}|C{2.5cm}}
        \toprule
        Parameter & Values \\
        \hline
        Hidden layer size & 100, 150, 200, 250, 300\textsuperscript{a}, 350, 400, 450, 500 \\
        Learning rate & 0.01, 0.0075, 0.005, 0.0025, 0.001\textsuperscript{a} \\
        Dropout & 0.2\textsuperscript{a}, 0.4  \\
        \bottomrule
    \end{tabular}
    \begin{minipage}{0.8\linewidth} 
        \footnotesize \centering
        \textsuperscript{a} Hyperparameters used for all trials in the manuscript.
    \end{minipage}
    \label{tab:SI-hyperparameter-grid-search}
\end{table}

\FloatBarrier

\subsection{Random noise injection} 
\label{subsec:random_noise_injection}

We used the normal distribution function of Numpy's Random sub-library to generate random noise profiles, which we define here as 2D arrays of Gaussian-distributed random noise. To most transparently study how noise in the dataset influenced ML, we ensured the following properties of the noise injection: (i) The noise profile injected to a given spectrum in one ML trial was identical to that injected into the same spectrum during a different, independent ML trial; (ii) No two 2DES spectra received identical noise profiles. While looping over the Hamiltonian index to add noise to each spectra, we used the following procedural steps: 

\begin{enumerate}
    \item Call numpy.random.default\_rng() with a known seed determined from the system index.
    \item Generate a 3D NumPy array of random noise, of size $\omega_1 \times \omega_3 \times t_2$, such that the 2D noise profiles are independent across all 250 $t_2$ time points.
\end{enumerate}

\noindent Seeding the NumPy RNG with Hamiltonian-specific seeds ensured property (i), while generating 3D NumPy arrays of noise is highly likely to have ensured (ii). The only variable between ML trials with differing amounts of noise injected was thus the standard deviation of the normal distribution ($\sigma$). 

\subsection{Signal-to-noise ratio}
\label{subsec:SNR-threshold}

\begin{figure}[h]
\centering
\includegraphics[scale=0.65]{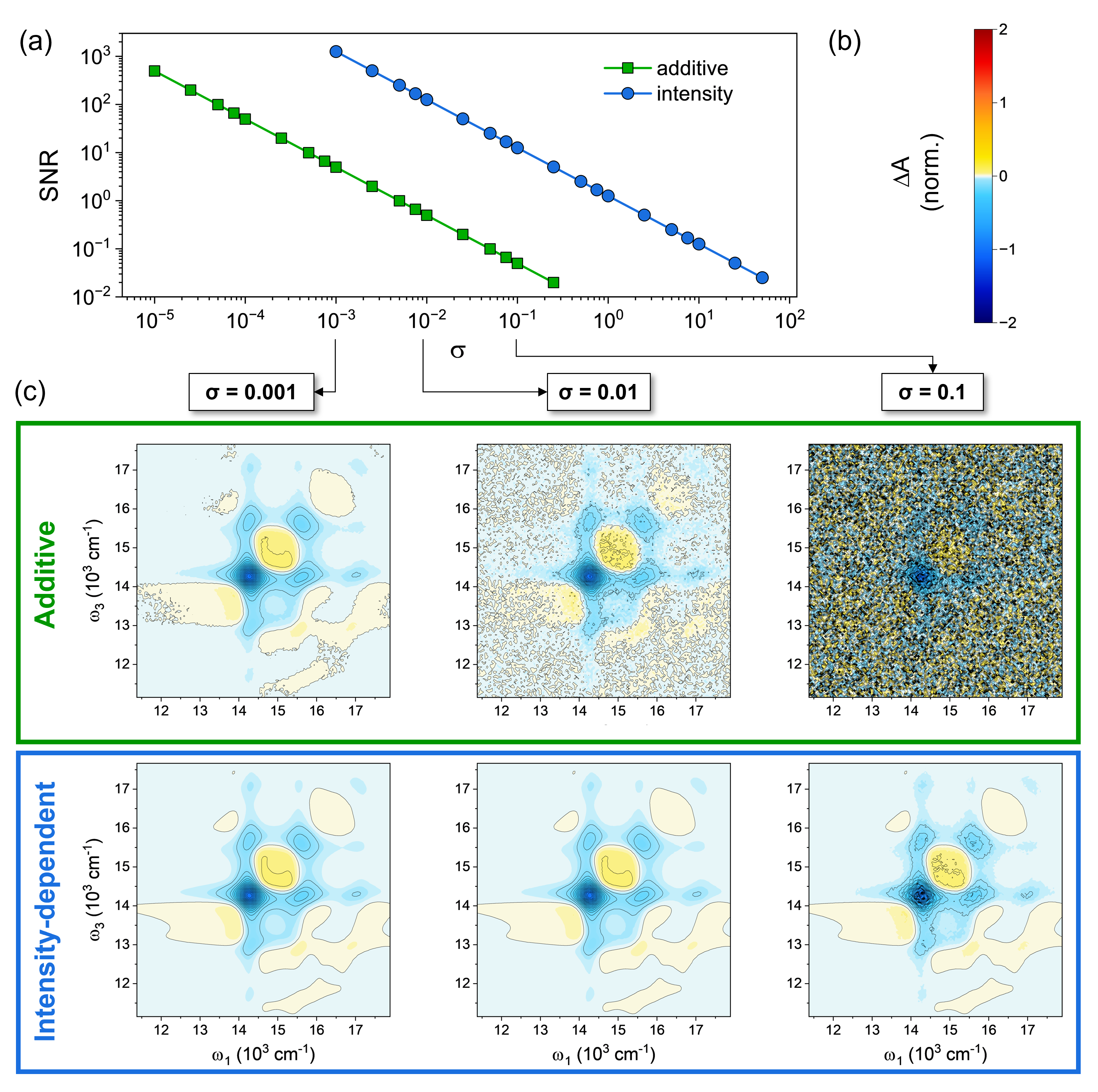}
\caption{(a) Values of the SNR (averaged over all spectra in the dataset) versus $\sigma$ for each noise category. (b) Scale bar for the 2D spectra in panel (c). (c) Representative spectra with additive (upper) and intensity-dependent (lower) noise for three $\sigma$ values.}
\label{fig:SI-SNR-comparison}
\end{figure}

The signal-to-noise ratio (SNR) is a common metric used in experimental spectroscopy. We define the SNR as

\begin{equation}
\label{eq:SNR}
    \mathrm{SNR} = \frac{\sigma_S}{\sigma + \alpha}
\end{equation}

\noindent where $\sigma_S$ and $\sigma$ are the signal- and noise-widths, respectively, and $\alpha$ is a constant ($10^{-10}$) to avoid division by zero. The value of $\sigma$ is the standard deviation of the Gaussian distribution used to generate random noise (\textit{vide supra}), while $\sigma_S$ for any given spectrum is the mean of the absolute-valued-spectrum. Figure \ref{fig:SI-SNR-comparison}a shows how the SNR depends on the category of noise.

\FloatBarrier

\begin{figure}[h]
\centering
\includegraphics[scale=0.35]{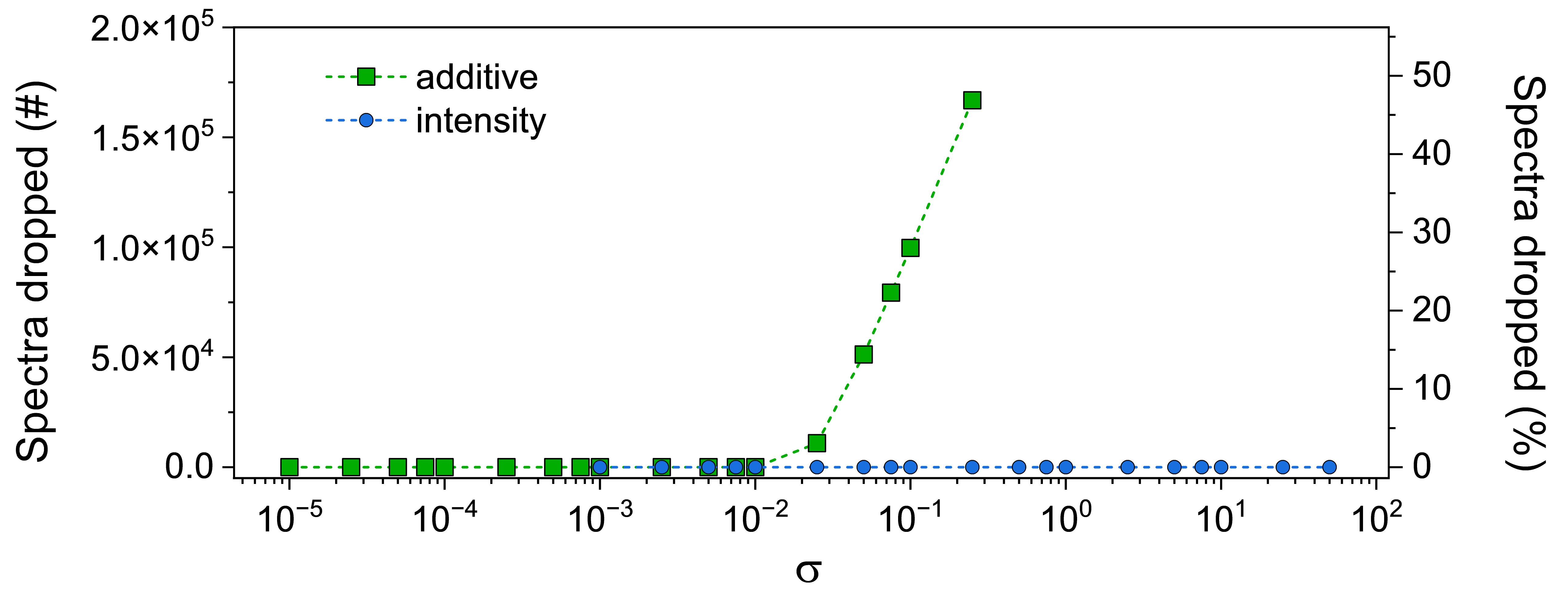}
\caption{Number of spectra removed from the dataset versus $\sigma$ for each noise category.}
\label{fig:cut-spectra}
\end{figure}

In practice, experimentalists do not attempt to interpret spectra that are saturated with noise beyond recognition. A similar practice should be incorporated into the training of NNs on noisy spectra. Thus, we defined a threshold SNR ($0.01$) such that spectra with SNR values below $0.01$ were removed from the training and testing datasets. Figure \ref{fig:cut-spectra} shows how the number of spectra removed from the full dataset as a function of $\sigma$. This threshold yielded no dropped spectra for trials with intensity-dependent noise.

\FloatBarrier

\subsection{Modeling the pump spectrum}

We modeled pump spectra in this work with a Gaussian profile,

\begin{equation}\label{eq:Gaussian_pulse}
    \Tilde{E}(\omega) = e^{\frac{-4log(2)(\omega-\omega_c)^2}{\Delta\omega^2}},
\end{equation}

\noindent where $\Tilde{E}(\omega)$ is the electric field as a function of the pump frequency ($\omega = \omega_1$), $\omega_c$ is the carrier frequency, and $\Delta\omega$ is the pulse bandwidth. Since convolution in the time-domain is equivalent to multiplication in the frequency domain,\cite{hamm2011concepts} we accounted for effects of the pump pulse spectrum by multiplying $R_{Abs}(\omega_1,t_2,\omega_3)$ by $\Tilde{E}(\omega_1)$.

\section{Additional results}
\label{sec:SI-additional-results}

\begin{figure}[h]
\centering
\includegraphics[scale=0.55]{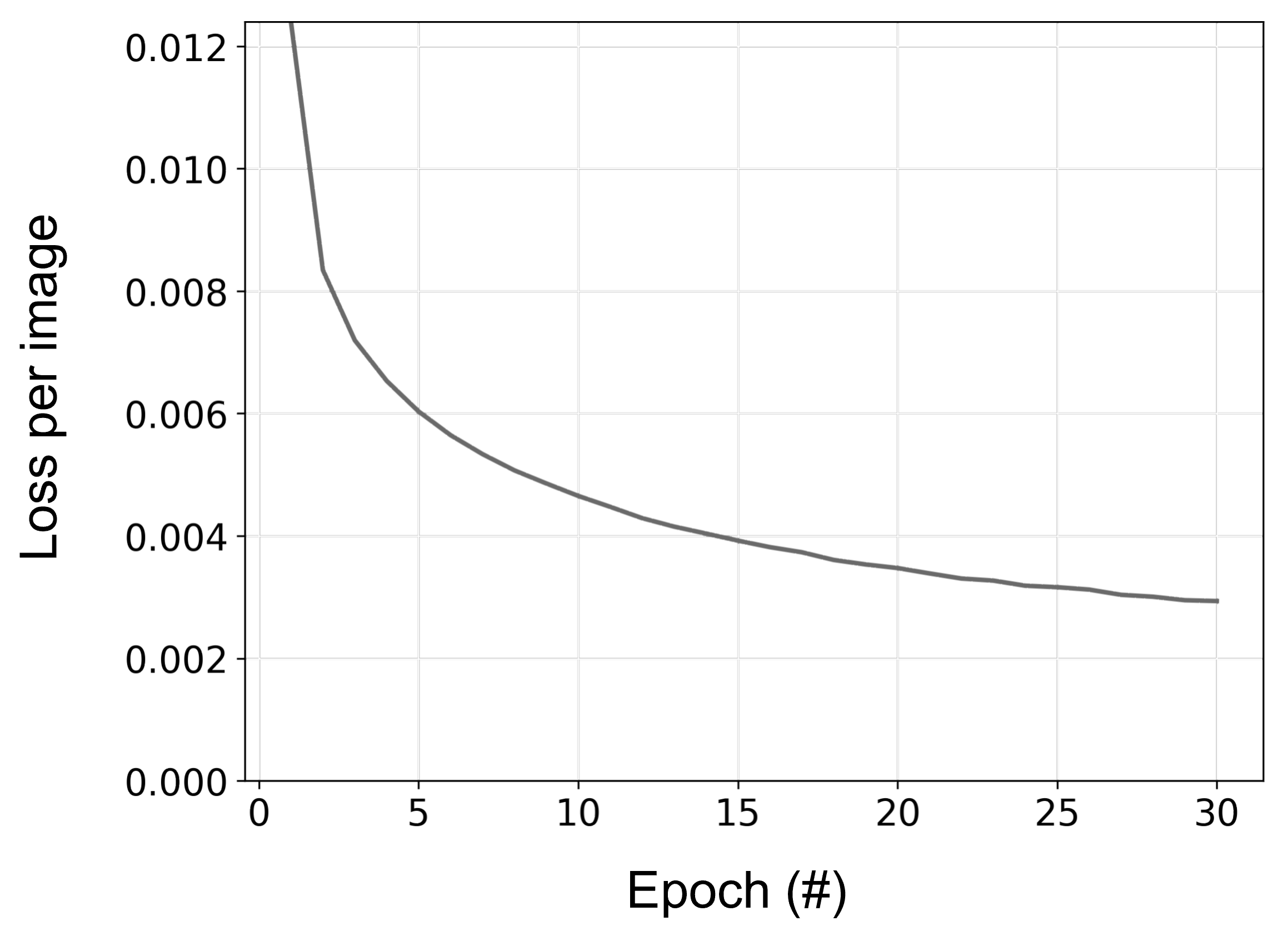}
\caption{Cross-entropy loss function for the un-polluted dataset.}
\label{fig:SI-training-loss}
\end{figure}

\begin{figure}[h]
\centering
\includegraphics[scale=0.57]{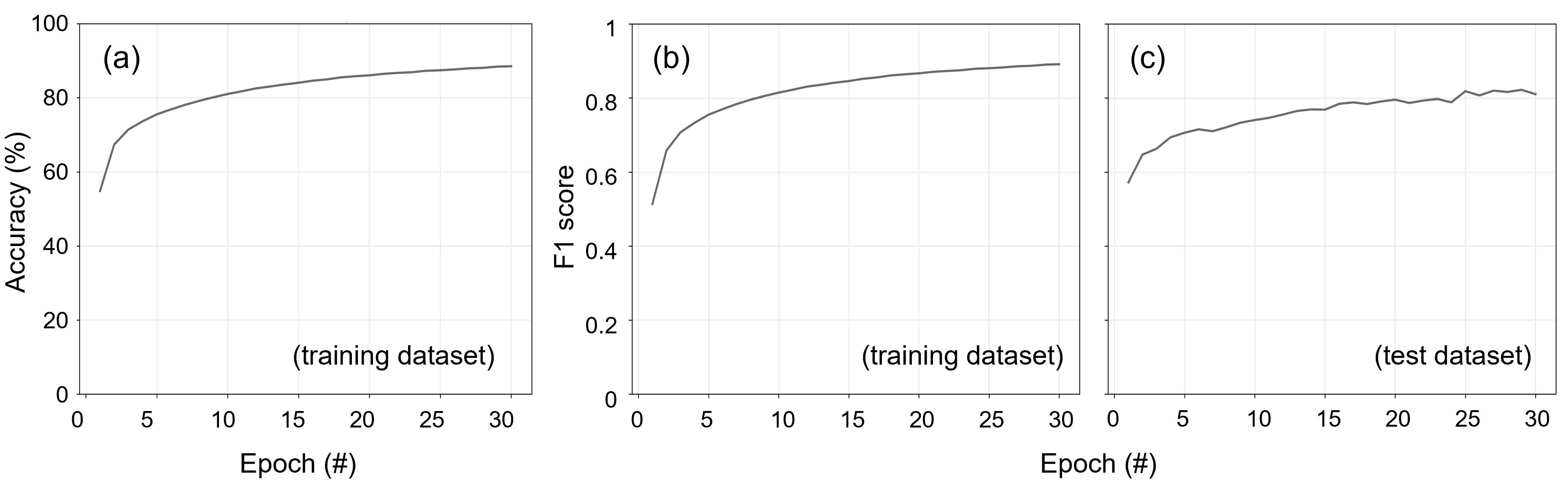}
\caption{Model performance vs. training epoch calculated as (a) accuracy and (b) F1 score (macro-averaged) on the clean (unpolluted) training dataset, and (c) F1 score (macro-averaged) on the clean (unpolluted) test dataset.}
\label{fig:SI-training-performance-clean}
\end{figure}

\begin{figure}[h]
\centering
\includegraphics[scale=0.7]{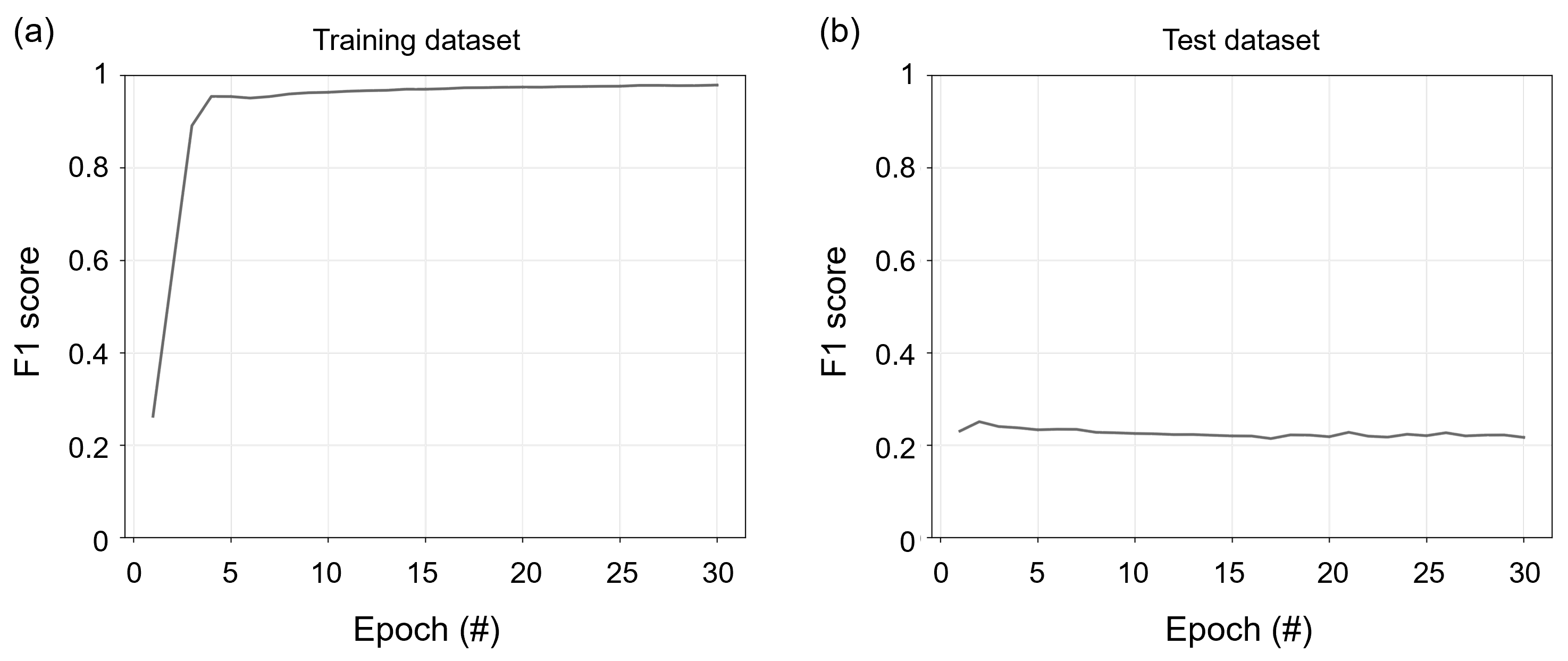}
\caption{Model performance (macro-averaged F1 score) vs. training epoch calculated on (a) training and (b) test datasets polluted with additive noise ($\sigma_{additive} = 0.25$).}
\label{fig:SI-training-performance-additive}
\end{figure}

Figures \ref{fig:SI-training-loss} through \ref{fig:SI-training-performance-additive} show analysis performed during the model training stage. For clean training and test datasets, the model performance exhibits the expected growth vs. epoch with an eventual plateau (Figure \ref{fig:SI-training-performance-clean}). In contrast, in the the ML trial with $\sigma_{additive} = 0.25$, the model performance exhibits clear signs of the model memorizing the noise signatures (Figure \ref{fig:SI-training-performance-additive}). Specifically, the model performance on the training dataset (Figure \ref{fig:SI-training-performance-additive}a) grows rapidly over the first five epochs while the test F1 score (Figure \ref{fig:SI-training-performance-additive}b) remains essentially invariant. Thus, all performance gains by the model on the training dataset vs. epoch are associated with memorizing the noise (as opposed to learning transferrable knowledge for the inverse classification problem at hand).  

\FloatBarrier

\begin{figure}[h]
\centering
\includegraphics[scale=0.59]{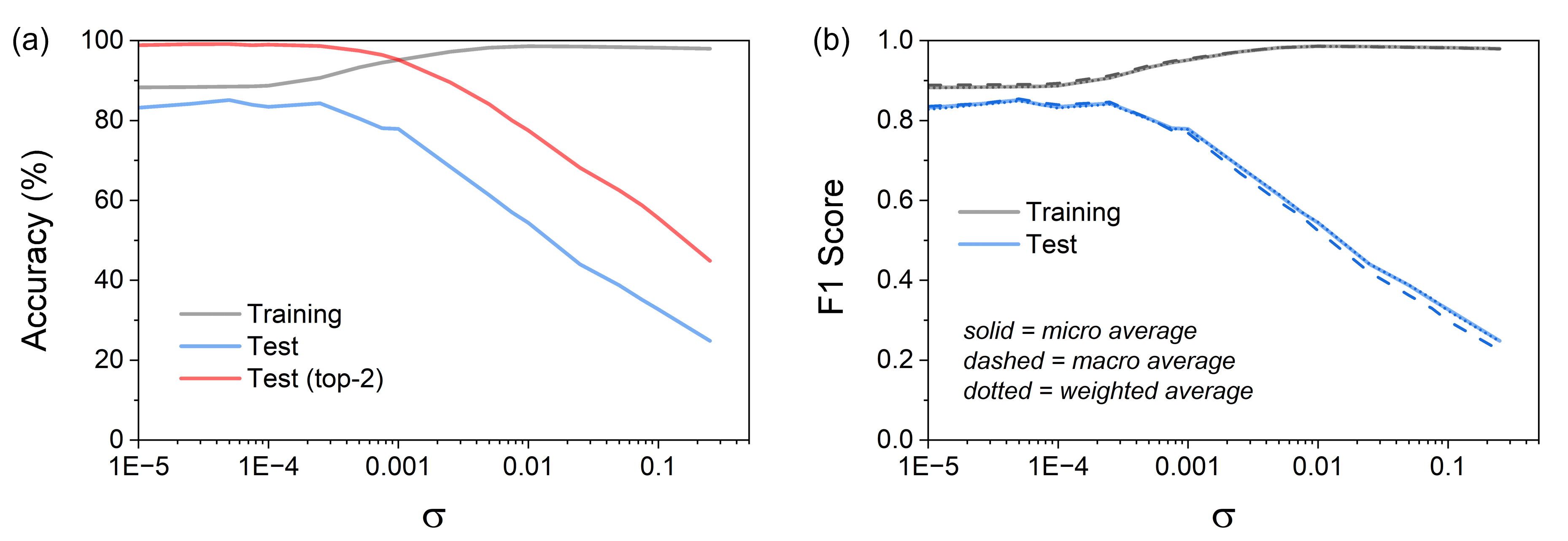}
\caption{Model (a) accuracy and (b) F1 score metrics for datasets with additive noise. In (b), the micro-, macro-, and weighted-averaged F1 scores are indicated by solid, dashed, and dotted lines, respectively.}
\label{fig:SI-performance-metrics}
\end{figure}

Figure \ref{fig:SI-performance-metrics} shows several model performance metrics on the training and test datasets as a function of additive noise $\sigma_{additive}$. For all datasets and pollutant types, we generally observed similar values for the accuracy and F1 scores (note line overlaps in Figure \ref{fig:SI-performance-metrics}b).

\begin{figure}[h]
\centering
\includegraphics[scale=0.6]{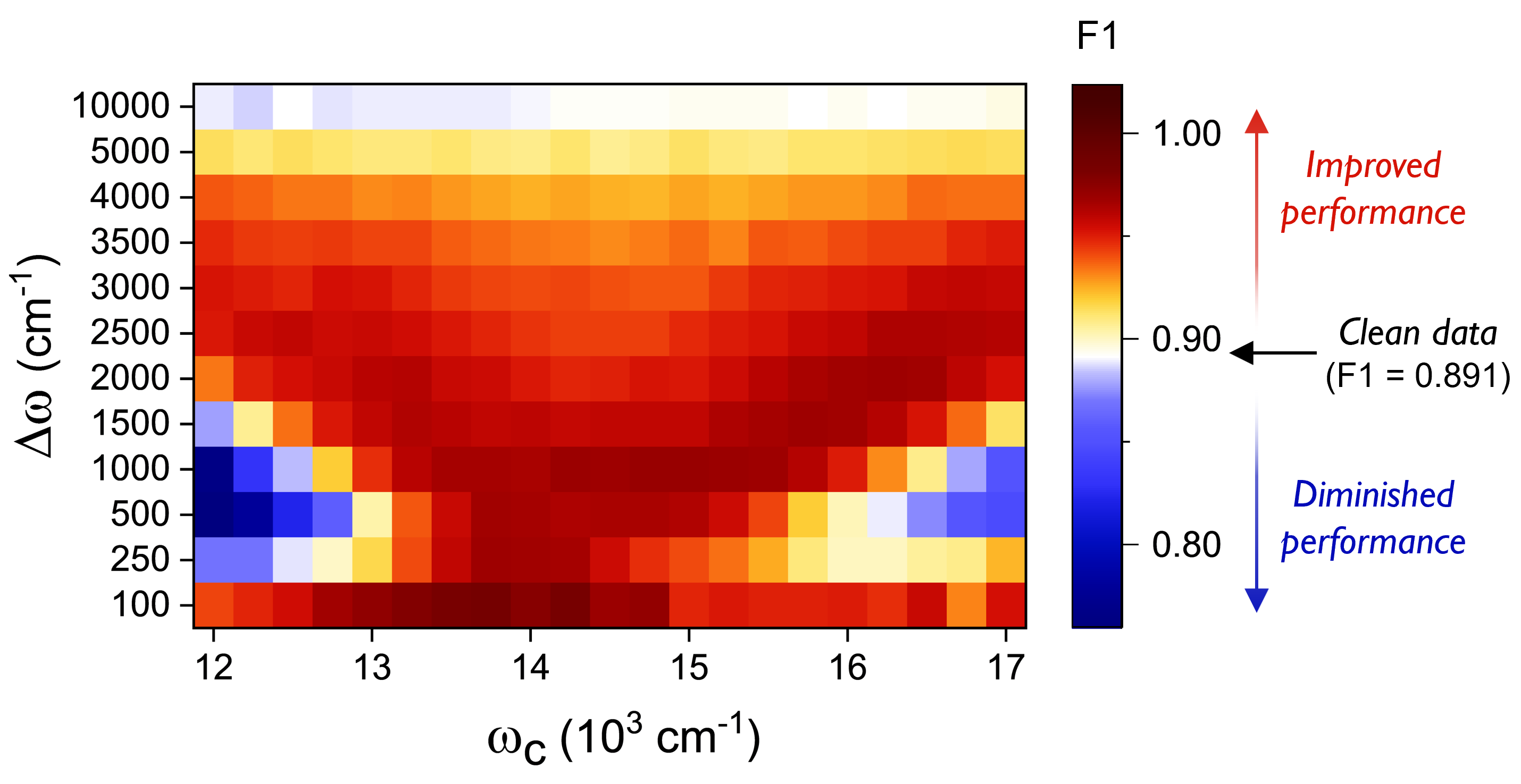}
\caption{Model F1 score for the training dataset as a function of $\Delta\omega$ and $\omega_c$ of the pump pulses. The color scale is based on the F1 score of $0.89129$ from the clean dataset.}
\label{fig:SI-dual-pump-training}
\end{figure}

\FloatBarrier

\twocolumn

\end{document}